\documentclass[review]{elsarticle}  	
\usepackage{comment}                		
\usepackage{graphicx}				
\usepackage{amssymb}
\usepackage{caption}
\usepackage{color}
\usepackage{balance}
\usepackage{booktabs}
\usepackage{rotating}
\usepackage{url}
\usepackage{hyperref}
\usepackage{listings}
\usepackage{color}
\usepackage{lineno,hyperref}
\usepackage{array}
\usepackage{rotating}
\usepackage{multirow}
\modulolinenumbers[5]

\definecolor{dkgreen}{rgb}{0,0.6,0}
\definecolor{gray}{rgb}{0.5,0.5,0.5}
\definecolor{mauve}{rgb}{0.58,0,0.82}
\journal{Computer \& Security}

\newcolumntype{M}{>{\centering\arraybackslash}m{1.9cm}}
\newcolumntype{N}{>{\centering\arraybackslash}m{1cm}}
\newcolumntype{L}{>{\arraybackslash}m{3.3cm}}

\begin{document}

\begin{frontmatter}
\title{Android Inter-App Communication Threats and Detection Techniques}

\author[mnit]{Shweta Bhandari\corref{cor1}}
\ead{er.shwetabhandari@gmail.com}
\author[olab]{Wafa Ben Jaballah}
\author[mnit]{Vineeta Jain}
\author[mnit]{Vijay Laxmi}
\author[labri]{Akka Zemmari}
\author[mnit]{Manoj Singh Gaur}
\author[labri]{Mohamed Mosbah}
\author[padua]{Mauro Conti}
\address[mnit]{Malaviya National Institute of Technology Jaipur (MNIT Jaipur)}
\address[olab]{Orange Labs, Paris, France}
\address[labri]{LaBRI - University of Bordeaux, CNRS, 33405 Talence cedex, FRANCE}
\address[padua]{University of Padua, Italy}

\cortext[cor1]{Corresponding author. Mobile: +91-7597385348}

\begin{abstract}
	With the digital breakthrough, smart phones have become very essential component for many routine tasks like shopping, paying bills, transferring money, instant messaging, emails etc. 
	Mobile devices are very attractive attack surface for cyber thieves as they hold personal details (accounts, locations, contacts, photos) and have potential capabilities for eavesdropping (with cameras/microphone, wireless connections). Android, being the most popular, is the target of malicious hackers who are trying to use Android app as a tool to break into and control device. Android malware authors use many anti-analysis techniques to hide from analysis tools. Academic researchers and commercial anti-malware companies are putting great effort to detect such malicious apps. They are making use of the combinations of static, dynamic and behavior based analysis techniques. 
	
	Despite of all the security mechanisms provided by Android, apps can carry out malicious actions through inter-app communication. One such inter-app communication threats is collusion. In collusion malicious functionality is divided across multiple apps. Each participating app accomplish its part and communicate information to another app through Inter Component Communication~(ICC).
	ICC does not require any special permissions. Also there is no compulsion to inform user about the communication. Each participating app needs to request a minimal set of privileges, which may make it appear benign to current state-of-the-art techniques that analyze one app at a time. 
	
	There are many surveys on app analysis techniques in Android; however they focus on single-app analysis. This survey highlights several inter-app communication threats, in particular collusion among multiple-apps. In this paper, we present Android vulnerabilities that may be exploited for carrying privilege escalation attacks, privacy leakage and collusion attacks. We cover the existing threat analysis, scenarios, and a detailed comparison of tools for intra and inter-app analysis. To the best of our knowledge this is the first survey on inter-app communication threats, app collusion and state-of-the-art detection tools in Android.
\end{abstract}
\end{frontmatter}

\section{Introduction}
\label{introduction}

\par Nowadays, mobile devices such as smartphones, are widely used for social networking, online shopping,  banking, etc. Mobile applications are increasingly playing an essential role in our daily life, making the safety guards in mobile operating systems an important concern for researchers and practitioners. Android is the most popular mobile operating system, with \textit{84\%} of the worldwide smartphone sales to end users in  first quarter of \textit{2016}~\cite{marketshare}, and over \textit{50} billion app downloads so far. The large popularity of Android and its open nature made it a primary target of hackers who are now developing malicious apps at an industrial scale~\cite{enck2014taintdroid, li2015apkcombiner, KElishICCMap15, dietz2011quire, lu2012chex, bugiel2011practical,shekhar2012adsplit, enck2011defending}.

An Android app consists of components and uses a special interaction mechanism to perform Inter-Component Communication (ICC). ICC enables modular design and reuse of functionality across apps and app components. In Android, ICC communication model is implemented as a message-passing system, where messages are encapsulated as Intent objects. Through Intents, an app (or app component) can utilize functionality exposed by another app (or app component), e.g. by passing a message to the browser to render content or to a navigation app to display a location and provide directions to it. This light communication model has been used by developers to design rich application scenarios by reusing existing functionality. Unfortunately, because many Android developers have limited expertise in security, the ICC mechanism has brought a number of vulnerabilities~\cite{chin2011analyzing, enck2014taintdroid, li2015apkcombiner, schlegel2011soundcomber, Felt2011, SbirleaPermissionFlow15}. Some of the ICC vulnerabilities viz. Activity hijacking vulnerability (where a malicious Activity is launched in place of the intended Activity), Intent spoofing vulnerability (where a malicious app sends Intents to an exported component which originally does not expect Intents from that app) etc. 

Different research efforts have investigated weaknesses from various perspectives~\cite{ravitch2014multi, li2015iccta, Sadeghi15Covert, MR-Droid, collusiveDataleak, collusiveRanking, intersection}, including detection of information leaks, analysis of the least-privilege principle, and enhancements to Android protection mechanisms.
Despite the significant progress, such security techniques are substantially intended to detect and mitigate vulnerabilities in a single app~\cite{bhandari2015draco,wei2014amandroid,chin2011analyzing, drebin, gordon2015information}, but fail to identify vulnerabilities that arise due to the interaction of multiple apps. Vulnerabilities due to the interaction of multiple apps, such as collusion attacks and privilege escalation chaining, cannot be detected by techniques that analyze a single app in isolation. Thus, there is a pressing need for security analysis techniques in such rapidly growing domains to take into account such communication vulnerabilities.

The principle of malware collusion has been recently described in a few research papers~\cite{felt2011survey,bugiel2011xmandroid, schlegel2011soundcomber, marforio2012analysis, markmann2013quantdroid, fang2014permission, covert, collusiveRanking, collusiveDataleak, MR-Droid} as the next step that malware writers may evolve into. Collusion refers to the scenario where two or more applications possibly (not necessary) developed by the same developer, interact with each other to perform malicious tasks. The danger of malware collusion is that each colluding malware only needs to request a minimal set of privileges, which may make it appear benign under single-app analysis mechanisms~\cite{KElishICCMap15, MR-Droid, collusiveDataleak, covert}.
The scenario could be think of as two utility apps one for cab booking and another is a browser app. Now cab booking app needs to access client's location and browser app needs to connect with the internet. Lets assume that both the apps are developed by same adversary and he intentionally puts a communication channel between these two apps. Whenever user invokes cab booking app, along with serving to the user it also sends location information to the browser app. Since browser have the access to internet, it can easily send the location information of the user to any command and control (C\&C) server. Malware writers have strong incentives to write colluding malware. 

{\color{blue}{
The wide usage of ICC calls in benign app pairs make accurate classification quite challenging~\cite{KElishICCMap15, primo, collusiveDataleak}. Academia and industry researchers have proposed solutions and frameworks to analyze, and detect the collusion attacks~\cite{bugiel2011xmandroid, ravitch2014multi, li2015iccta, li2015apkcombiner, covert, collusiveDataleak, MR-Droid, collusiveRanking, intersection, bhandari2017poster, klieber2014android}. Some of these are even available as open-source as~\cite{li2015iccta, li2015apkcombiner, klieber2014android, collusiveDataleak}. The solutions can be characterized using three broad types of analysis: Static analysis, dynamic analysis and policy enforcement based analysis. 

In~\cite{collusiveDataleak}, authors propose a tool named DIALDroid, as the most recent state-of-the-art inter-app ICC analysis tool for large scale detection of collusion and privilege escalation. They also provide the first inter-app collusion real-apps benchmark of $30$ apps. Till now, this is the most efficient tool available in the literature for inter-app vulnerability detection. MR-Droid~\cite{MR-Droid} aims to detect inter-app communication threats specifically intent hijacking, intent spoofing and collusion. It proposes a MapReduce based framework to scale up compositional app analysis. DidFail~\cite{klieber2014android} is another state-of-the art to detect intra-component and inter-component information flow in a set of apps. 
In~\cite{bugiel2011xmandroid}, authors propose XMandroid, that is the first approach for detecting collusion attacks in Android platforms. It claims to identify privilege escalation in case of pending intents and transmission channels between dynamically built components such as broadcast receivers.
FUSE~\cite{ravitch2014multi} is a tool that starts by single-app static analysis accompanied with lint tool to mitigate limitations of static analysis followed by multi-app information flow analysis.
IccTA~\cite{li2015iccta} is a static taint analyzer to detect privacy leaks between components in Android apps. If combined with APKCombiner~\cite{li2015apkcombiner}, it can also detect inter-app leakage paths.
}}

This survey paper aims to present a general review about inter-app communication threats in particular, collusion attacks in Android framework. It provides a better understanding of the key research challenges. We present an abstract definition of collusion and highlight its origin. Along the way, we cover the Android model, the communication and permission model of Android and the main vulnerabilities that lead to a possible collusion attack. We also cover the existing threat analysis and a detailed comparison of techniques for intra and inter-app analysis. This review gives an insight into the strengths and shortcomings of the known tools and provides a clear comparison for the researchers between these tools. Finally, we present an insight into our future research directions. 

This survey paper is organized as follows. Section~\ref{androidDevelopmentModel} presents Android model. In Section~\ref{communicationModel}, we present the Inter Process Communication (IPC) model as one of the key features of programming model in Android. Then, in Section~\ref{vulnerabilityAttack}, we present Android security risks. In Section~\ref{examples}, we elaborate collusion by providing a formal definition and cases where collusion attack is possible followed by the main challenges to detect collusion attack. In Section~\ref{interAppAnalysis}, we review the inter application analysis. Section~\ref{stateOfArt} recalls state-of-art approaches, a thorough comparison between them for collusion detection and lessons learned. In Section~\ref{conclusion}, we conclude the paper and we present an insight into our future research directions. 

\section{Android}
\label{androidDevelopmentModel}
Android is developed under the Android Open Source Project (AOSP), promoted by the Open Handset Alliance (OHA) and maintained by Google~\cite{androidwiki}. Android is developed on top of Linux kernel due to its robust driver model, 
efficient memory,  process management, and networking support for the core services. Linux Kernel is customized specifically for the embedded environment consisting of limited resources. 

Android apps are written in java; however, the native code and shared libraries are developed in C/C++ to support high performance~\cite{androidndk}. There are two runtime environments available in Android viz. Dalvik Virtual Machine (DVM) and Android Runtime (ART). In DVM, dex file of the Android apps are translated to their respective native representations on demand using just-in-time (JIT) compiler. However, in case of ART, ahead-of-time (AOT) compilation is performed i.e. at the time of installation itself apps are compiled to a ready-to-run state~\cite{dvmart}. Therefore, ART massively improves the performance and battery life of Android device. 

Once the OS boot completes, a process known as zygote (parent of all apps) initializes. As zygote starts, it preloads all necessary Java classes and resources, starts System Server and opens a socket /dev/socket/zygote to listen for requests for starting applications. Thus zygote process expedites the app launching process. 

In the following, we present the main Android app composition followed by Android security model viz. application signing, application permission, and sandboxed environment. 

\subsection{Android App Composition}
Android applications are distributed as binaries in a regular format based on zip files with .apk as file extension. It usually contains the following files and directories~\cite{apkwiki}.
\begin{enumerate}
\item \texttt{Manifest file:} Manifest file is an XML configuration file (AndroidManifest.xml) one per app. It is used to declare various components of an application, their encapsulation (public or private) and the permissions required by the app. Android APIs offer programmatic access to mobile device-specific features such as the GPS, vibrator, address book, data connection, calling, SMS, camera, etc. These APIs are usually protected by permissions. For example the Vibrator class, to use the \texttt{android.os.Vibrator.vibrate(long milliseconds)} function, which starts the phone vibrator for a number of milliseconds. The permission \texttt{android.permission.VIBRATE} must be declared in the app manifest file.  
\item  \texttt{dex file:} A Dalvik executable (classes.dex), which contains the bytecode of the program.
\item \texttt{res directory:} Resources including string literals, their translations, and references to binary resources.
\item \texttt{layout directory:} XML layouts describing user interface elements.
\item \texttt{lib directory:} The directory containing the compiled code that is specific to a software layer of a processor.
\item \texttt{assets directory:} The directory containing applications assets, which can be retrieved by AssetManager.
\end{enumerate}
An android app is composed of any combination of the following four components:
\begin{itemize}
	\item Activities: The Android libraries include a set of GUI components specifically built for the interfaces of mobile devices, which have small screens and low power consumption. One type of such component is Activities, that represent screens which are visible to the user;
	\item Services: They perform background computation;
	\item Content Providers: They act as database-like data stores;
	\item Broadcast Receivers: They handle notifications sent to multiple targets.
\end{itemize}


%

\subsection{Android Security Model}
\label{appsecuritymodel}
Android security depends on restricting apps by combining app signing, sandboxing, and permissions.
\subsubsection{App Signing}
App signing is a prerequisite for inclusion in the official Android market (Google Play Store). App signature is the point of trust between Google and the third party developers to ensure app integrity and the developer reputation. Most developers use self-signed certificates that they can generate themselves, which do not imply any validation of the identity of the developer. Instead, they enable seamless updates to applications and enable data reuse among sibling apps created by the same developer~\cite{signing}.

\subsubsection{App Permission}
App permission model regulates how applications access certain sensitive resources, such as users' personal information or sensor data (e.g., camera, GPS, etc.).  For instance, an application must have the
$READ\_CONTACTS$ permission in order to read entries in a user's phone \cite{Felt:2011:APD:2046707.2046779}. System permissions are divided into four protection levels. The two most relevant levels to this manuscript are normal and dangerous permissions. Normal permissions require when the app needs to access data or resources outside the app's sandbox, but involves very little risk to the user's privacy or the operation of other apps. For example, permission to set the alarm is a normal permission. Dangerous permissions are required when the app wants data or resources that involve user's private information or could potentially affect user's stored data or the operation of other apps. For example, the ability to read user's contacts is a dangerous permission~\cite{appperm}. Applications can also define their own permissions in order to restrict the use of components in an application that can perform sensitive tasks. The third level of permission is Signature permission that is used by the developers to transfer resources and data between their own applications meanwhile safeguarding them against applications of other developers~\cite{Egners:2012:MAP:2360018.2360209}. Lastly, SignatureOrSystem permission which is high-level permission that includes changing security settings, installing an application, etc. These permissions are maintained by OS developers and manufacturers. They are granted by System to those applications which are either contained in the system image or signed by the same certificate as of system image~\cite{barrera2010methodology}.
\par App permissions play important role in malware detection. There exist rich literature embodies  tools that attempt to identify malicious applications through their permission requests~\cite{felt2011permission, Felt2011, Au:2012:PAA:2382196.2382222}. Researchers have also developed static and dynamic analysis tools to analyze Android permission specifications \cite{Egners:2012:MAP:2360018.2360209,sbirlea2013automatic, barrera2010methodology}. Permission enforcement techniques are also proposed \cite{Felt:2011:APD:2046707.2046779}.

\subsubsection{Sandboxed Environment}
Android apps are executed in a sandboxed environment to protect the system, the user data, the developer apps, the device,  the network, and the hosted applications, from malware~\cite{ Egners:2012:MAP:2360018.2360209}. Each app process is protected with an assigned unique id (UID) within an isolated sandbox. The sandboxing restrains other apps or their system services from interfering the app~\cite{rasthofer2014droidforce}.
Android protects network access by implementing a feature Paranoid Network Security, a feature to control Wi-Fi, Bluetooth and Internet access within the groups. If an app has permission for a network resource (e.g., Bluetooth), the app process is assigned to the corresponding network access id. Thus, apart from UID, a process may be assigned one or more group id (GIDs).
An app must contain a PKI certificate signed with the developer key. App signing procedure places an app into an isolated sandbox assigning it an unique UID. If the certificate of an app A matches with an already installed app B on the device, 
Android assigns the same UID (i.e., sandbox) to apps A and B, permitting them to share their private files and the manifest defined permissions. This unintended sharing can be exploited by the malware writers as naive developers may generate two certificates. It is advisable for the developers to keep their certificates private to avoid their misuse. The Android sandbox relies on, and augments, the Linux kernel  isolation facilities. While sandboxing is a central security feature, 
it comes at the expense of interoperability. In many common situations, apps require the ability to interact. For example, the browser app should be capable of launching the Google Play app if the user points toward the Google Play website~\cite{appsandbox}.

\section{Inter-Component Communication}
\label{communicationModel}
Inter Process Communication (IPC) is known as Inter Component Communication (ICC) in Android~\cite{enck2009understanding}. It is the key features of Android programming model. It allows a component of an application to access user's data and can transfer it to another component of same or other application within the same device, or to an external server. 
ICC helps to eliminate duplication of functionality in different applications. 
Developers can leverage data and services provided by other applications. For example, a cab booking application can ask Google Maps for client's or driver's location information. This communication between applications can reduce developer's burden and facilitate functionality reuse.

To support inter-component communication, there exists conventional methods called \textit{overt channels} and non-conventional methods called \textit{covert channels}.
Covert channels are intentionally used to hide the messages or communication. Any app using covert channel as a medium of communication can be suspected as malicious. Although overt channels are perfectly benign and widely used for communication in Android apps. The main focus of this paper is to show that how a set of apps appear perfectly benign can carry out a threat. Therefore in this section, we will elaborate overt channels and provide glimpse of covert channels.
\subsection {Overt channel}
\textit{Overt channel} is an unconcealed medium provided by Android framework for communication. In the following, we present the main channels that come under this such as Intents, Content Providers, Shared Preferences, External Storage, and Remote Method Calls. 

\subsubsection{Intents} 
Intents enable components of an application to invoke other components of the same or different applications. It is also used to pass data between different components through Bundles. It optionally contains destination component name or action string, category and data. Intents are the preferred message passing mechanism for asynchronous IPC in Android. The Android API defines methods called ICC methods that can accept intents and perform actions accordingly. For example, \texttt{startActivity(Intent)}, \texttt{startService(Intent)} etc.
\vspace{2mm}
\par ICC is widely facilitated through Intents. In ~\cite{li2015iccta}, authors highlighted that \textit{2955}/\textit{33258} applications use ICC through intents. Intents can be sent to three out of four components. Based on destination of ICC calls, they are categorized into two broad categories:

\subsubsection*{Implicit Intent} 
Implicit intents are used when the receiver of the intent is not fixed~\cite{li2015iccta}. Whenever an app wants to send the intent to all the registered components (registration is done using an intent filter in the manifest file) within and across the installed apps. When an app invokes API call with implicit intent then depending on the type of component, framework serves the calls.
\begin{itemize}
	\item If the calls  are intended for  activity, then  users are asked for the choice.
	\item If the calls are intended for service, then the framework will randomly choose one of the registered services.
	\item If the calls are intended for broadcast receivers, then the framework delivers to all the receivers.
\end{itemize}
   In the following, we present a sample code of implicit intent where \texttt{"com.example.msgSendFirst"} is the action string:
\begin{lstlisting}
 /**
  * Implicit Intent
  */
Intent intent = new Intent("com.exampke.msgSendFirst");
startActivity(intent);
\end{lstlisting}

\subsubsection*{Explicit Intent} 
Explicit intents are used when receiver of the intent is fixed. When an app invokes API call with explicit intent, the framework will deliver the intent to the component that is mentioned in the intent. A Sample code of explicit intent where \texttt{"this, LoginActivity.class"} is the address of the destination component:
\begin{lstlisting}
 /**
  * Explicit Intent
  */
Intent intent = new Intent(this,LoginActivity.class);
startActivity(intent);
\end{lstlisting}

\subsubsection{Content Provider}
Content Providers are used to transfer structured data across components of same or different apps. It stores information in tables like relational databases. To access or modify data in Content Provider, apps need \texttt{ContentResolver} objects. An app can also attach read and write permissions to the content provider it owns.
\begin{lstlisting}
public class SchoolProvider extends ContentProvider {
  /**
  * Declaration
  */
   static final String PROVIDER_NAME = "com.example.provider.School";
   static final String URL = "content://" + PROVIDER_NAME + "/students";
   static final Uri CONTENT_URI = Uri.parse(URL);
   ...........
   ...........
   
   /**
   * Database specific constant declarations
   */
   private SQLiteDatabase db;
   static final String DATABASE_NAME = "School";
   static final String STUDENTS_TABLE_NAME = "students";
   .............
   ..............
   
   /**
   * Helper class that actually creates and manages 
   * the provider's underlying data repository.
   */
   private static class DatabaseHelper extends SQLiteOpenHelper {
      DatabaseHelper(Context context){
         super(context, DATABASE_NAME, null, DATABASE_VERSION);
      }
      
      @Override
      public void onCreate(SQLiteDatabase db)
      {
         db.execSQL(CREATE_DB_TABLE);
      }
      public boolean onCreate() {
      /**
      * Create a write able database which will trigger its 
      * creation if it doesn't already exist.
      */
      }
       public Uri insert(Uri uri, ContentValues values) {
      /**
      * Add a new student record
      */
      }
      public Cursor query(Uri uri, String[] projection, String selection,String[] selectionArgs, String sortOrder) {
      /**
      * Code for querying the database
      */
      }
      public int delete(Uri uri, String selection, String[] selectionArgs) {
      /**
      * Code for deleting records from the database
      */
      public int update(Uri uri, ContentValues values, String selection, String[] selectionArgs) {
      /**
      * Code for updating records from the database
      */
      }
\end{lstlisting}

\subsubsection{Shared Preference}
Shared Preference is an operating system feature that allows apps to store key-value pairs of data. Its purpose is to be used to store preferences information. Apps can use key-value pairs to exchange information if proper permissions are defined when accessing and storing data.
\begin{lstlisting}
      /**
      * Declaration
      */
      SharedPreferences sharedPref = getActivity().
      getPreferences(Context.MODE_PRIVATE);
      /**
      * Write to Shared Preferences
      */
      SharedPreferences.Editor editor = sharedPref.edit();
	  editor.putInt(getString(R.string.
	  saved_high_score), newHighScore);
      editor.commit();
      /**
      * Read from Shared Preferences
      */
      int defaultValue =  getResources().getInteger
      (R.string.saved_high_score_default);
      long highScore = sharedPref.getInt
      (getString(R.string.saved_high_score), 
      defaultValue);
\end{lstlisting}

\subsubsection{External Storage}
External Storage is the storage space external to an app. It includes USB connection, SD card and even non-removable storage. Apps accessing the external storage need to declare the \texttt{READ\_EXTERNAL\_STORAGE} permission. Apps declaring the \texttt{WRITE\_EXTERNAL\_STORAGE} can write and read from external storage.
\begin{lstlisting}
		/**
      	* Write to SD card
      	*/
		File myFile = new File("/sdcard/mysdfile.txt");
		myFile.createNewFile();
		FileOutputStream fOut = new FileOutputStream(myFile);
		OutputStreamWriter myOutWriter = new OutputStreamWriter(fOut);
		myOutWriter.append(txtData.getText());
		myOutWriter.close();
		fOut.close();
	    /**
        * Read from SD Card
        */
        File myFile = new File("/sdcard/mysdfile.txt");
		FileInputStream fIn = new FileInputStream(myFile);
		BufferedReader myReader = new BufferedReader(
					new InputStreamReader(fIn));
		myReader.close();
\end{lstlisting}

\subsubsection{Remote Method Calls}
Remote methods enable to make method calls that look local but are executed in another process. It is same as remote procedural calls (RPC) in other systems. In Android, due to sandboxing, one process cannot access the memory of another process. In order to communicate, they need to pack their objects into the primitives that operating system can understand and unpack them again at receiver's end. This is facilitated by Binder, Messenger and AIDL.
\begin{itemize}
\item \textit{Binder}: Binder supports remote method calls within same application without the support of multi-threading. Binders are the entity which allows activities and services to obtain a reference to another service. It allows not simply send messages to services but directly invoke methods on them. Binder class provides direct access to public methods in the service but can be used only when the service is used by the local application and in the same process. For example, it would work if a music application that needs to bind an activity to its own service that's playing music in the background.

\item \textit{Messenger}: Messenger supports remote method calls across the applications without the support of multi-threading. It represents a reference to a Handler that can be sent to a remote process via an Intent. Messenger provides an interface to the service to communicate with remote processes. This allows inter-process communication without the use of AIDL. This can be used in the case where remote IPC is required but multi-threading support by the service is not required.

\item \textit{AIDL}: AIDL supports remote method calls across the applications with the support of multi-threading. Android provides an API to handle marshalling and unmarshalling of objects called Android's Interface Definition Language (AIDL). AIDL is necessary only if remote access of the service is required for IPC and want to handle multithreading in that service. AIDL is complex to implement as this interface sends simultaneous requests to the service, which must then handle multi-threading.

\end{itemize}

\subsection{Covert channel}
\textit{Covert channel} is a secret medium which exploits shared resources and use them for communication~\cite{bugiel2011xmandroid}. Timing channels and storage channels come under the covert channel. 
\subsubsection{Timing channels}
In timing channel, the information between applications is synchronously transmitted using shared resource having no storage capability. Battery use, Phone call frequency are  examples of timing channels.
\subsubsection{Storage channels}
In storage channel, the information between  applications is asynchronously transmitted using shared resource with storage. Phone call logs, Content providers are examples of storage channels.

{\section{Android Security Risks and Consequences}
\label{vulnerabilityAttack}
Android ensures security through its sandbox model, application signing and the permission model for managing IPC effectively and efficiently. In spite of these measures, Android is vulnerable to many security risks. According to the recent OWASP mobile security report~\cite{OWASP_report}, out of \textit{91} reported security risks, \textit{85} are recorded to be present in Android. This makes Android security a serious concern.
These risks are outcome of either maliciously exploiting the legitimate procedures provided by android such as ICC,  or taking advantage of unchecked processes occurring in the system. In the following, we focus on Intent based attacks and their consequences.

\subsection{Intent based attacks}
\label{Intentattacks}
We focus our attention on the security challenges of Android communication from the perspectives of Intent sending and receiving.  In section~\ref{Intentspoofing}, we focus on the Intent receiving, and consider vulnerabilities related to receiving Intents coming from other applications. In Section~\ref{Intenthijacking}, we consider how sending Intents to the wrong application can leak user information. 

\subsubsection{Intent Spoofing}
\label{Intentspoofing}
Intent spoofing refers to a typical scenario where a vulnerable app has a component that expects Intent from Android framework or itself. If the component is exposed, then other malicious apps can send forged Intents, and then spoof this app in order to trigger 
misbehaved actions. In the following, we classify the Intent spoofing to three subclasses~\cite{chin2011analyzing}: malicious broadcast injection, malicious activity launch, and malicious service launch.

\paragraph{Malicious Broadcast Injection}
Broadcast receivers are vulnerable to malicious broadcast injection when they receive Intents with system actions~\cite{chin2011analyzing}.
An example  scenario of this attack is that some Intents 
contain action strings that only the operating system may add to broadcast Intents. 
If a malicious application sends an Intent explicitly addressed to the target Receiver, without containing the system action string. The Receiver will be tricked into performing functionality that only the system should be able to trigger if it does not check the Intent's action.

\paragraph{Malicious Activity Launch}

A malicious activity launch is executed by other applications with explicit or implicit Intents. The impact of this attack is that the malicious Activity's UI will load instead of the targeted one. In~\cite {chin2011analyzing}, 
the authors classify three types of possible attacks when launching malicious activity: \begin{itemize}
\item Modification of data in the background caused due to non verification of the origin of the Intent or leading to a change in the application state; 
\item A user can be misleaded between malicious and victim applications. She might make changes to the victim application while believing she is interacting with the malicious one.
\item A victim application could leak some sensitive information returning a result to its caller upon completion. 
\end{itemize}

\paragraph{Malicious Service Launch}
In the same way as an Activity, a Service, if not protected with permissions, any application can bind it. This vulnerability could lead even to leak information or perform unauthorized tasks, depending on the type of Service~\cite{chin2011analyzing}. 

\subsubsection{Intent Hijacking}
\label{Intenthijacking}
The Intent hijacking threat is illustrated when an Intent could not reach the intended recipient via an implicit ICC, and then it may be hijacked by an unauthorized app~\cite{chin2011analyzing}. We classify this threat based on the type of the sending component: broadcast receivers hijacking, activity hijacking, and service hijacking.

\paragraph{Broadcast Receivers Hijacking}
Broadcast Receivers can be vulnerable to active denial of service attacks or eavesdropping. An eavesdropper can read the contents of a broadcast Intent without interrupting the broadcast. This is a risk whenever an application sends a public broadcast. A malicious Broadcast Receiver could eavesdrop on all public broadcasts from all applications by creating an Intent filter that lists all possible actions, data, and categories~\cite{chin2011analyzing}. 
\paragraph{Activity Hijacking}
In an Activity hijacking attack, a malicious Activity is launched instead of the intended Activity~\cite{lu2012chex}. This attack works as follows: 
The malicious Activity registers to receive another applications implicit Intents, and it is then started instead of the expected Activity.  The impact of this attack is dreadful since the malicious activity could read the data in the Intent and then immediately relay it to a legitimate activity~\cite{chen2014peeking} into the victim application.
\paragraph{Service Hijacking}
The service hijacking attack occurs when a malicious service intercepts an Intent designed for a legitimate service~\cite{li2014automatically}. 
The impact of this attack is that the initiating application establishes a connection with a malicious service instead of the legitimate one. The malicious service can steal data and lie about completing requested actions~\cite{kantola2012reducing}.

\subsection{Android Risk Consequences}
In the following, we focus on two major consequences of the Android security risks: privilege escalation, and privacy leaks. 

\subsubsection{Privilege Escalation}
Android security framework enforces permission protected model that allows the user to regulate the access of data by an application. It has been shown that applications can bypass this security model by exploiting transitive permission usage known as privilege escalation~\cite{heuser2016droidauditor},~\cite{markmann2013quantdroid},~\cite{bugiel2011xmandroid},~\cite{bugiel2012towards}. This refers to the scenario where two or more applications with a limited set of permissions
communicate with each other to gain indirect privilege
escalation and can perform unauthorized actions. \\
Figure 1 shows an example of privilege escalation attack. In this example, an android device contain $2$ apps - \texttt{Contacts manager} and \texttt{News}. Contact manager includes permissions \texttt{READ\_CONTACTS} and \texttt{WRITE\_CONTACTS}. On the other hand, News app includes permission \texttt{INTERNET}. Contact manager cannot access internet and news app cannot access contacts stored on the device. However, they can communicate via Intents. Contact manager sends an Intent carrying contact information as a payload to news app. As news app can access internet, it can transmit it to outside world. Hence, news app can access contacts, even if it does not contain permission to do so. Hence its privileges got escalated.
\begin{figure}
	\label{motivationEg}
	\centering
	\includegraphics[width=8cm,height=10cm,keepaspectratio]{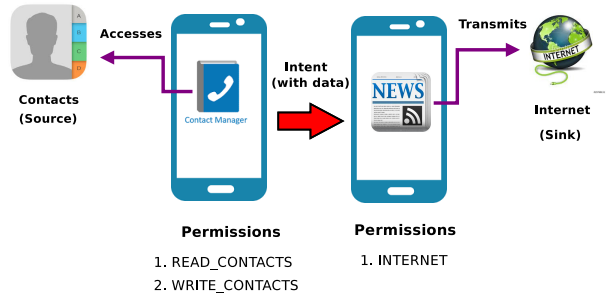}
	\caption{Privileged Escalation attack scenario}
\end{figure}
\subsubsection{Privacy Leaks}
\label{privacyLeaks}
Privacy leak occurs if there is a secret (without user consent) path from sensitive data as source to statements sending this data outside the application or device, called sink. This path may be within a single component or across multiple components. Thus, analyzing components separately is not enough to detect leaks. It is necessary to perform an inter-component analysis of applications. Android app analysts could leverage such tools to identify malicious apps that leak private data. For the tool to be useful, it has to be highly precise and minimize the false positive rate when reporting applications leaking private data. For example, IccTA, an inter-component communication Taint Analysis tool~\cite{li2015iccta}. It is for a sound and precise detection of ICC links and leaks.\\ 
\par Recent works have demonstrated that Android apps exhibit different privacy leaks, that are mainly build around the collusion attack~\cite{enck2011study, bartel2012automatically, suarez2014evolution,armando2013bring}. The main 
vulnerability comes from the fact that  these leaks are exacerbated by several applications that can interact to leak data using the inter-app communication mechanism~\cite{enck2014taintdroid}. The aforementioned security risk could lead to the collusion attack resulting in privacy abuse. Through the inter-app ICCs, two or more apps can collude to perform malicious actions. We give more details about the collusion attacks in Section~\ref{examples}.

\section{Collusion}
\label{examples}
The Android security model is designed to protect data, applications and devices from security threats. It is guarding apps by combining app signing, sandboxing, and permissions. Unfortunately, these restrictions can be bypassed by colluding apps. The combined permission of these apps allow them to carry  out attack, that could not be possible by a single app.  Let us consider the following example where a collusion consists of one app permitted to access some personal data, and this app passes the data to a second app that is allowed to transmit the data. Moreover, the Android OS does not check if an app that is accessing a permission-protected resource through another app has itself requested that permission. We believe that collusion is worth investigation since it could be exploited by criminals, and become a serious threat in the near future.
\par In this section, we start with a brief history of the origin of collusion. Then we define collusion along with its categories on the grounds of application properties. We also highlight some challenges faced in detecting collusion.  
For the sake of clarity, we define some terminologies such as \textit{Sensitive Resource Access:} When an Android app access the system resource that is protected with some dangerous permission then that access to the resource is called Sensitive Resource Access. \textit{Sensitive Information:} Any piece of data that generates from sensitive resource access becomes Sensitive Information. \textit{Leakage:} It happens when the sensitive information moves out of device boundaries without user consent.

\subsection{History}
The problem of colluding apps can be traced back to confused deputy attack. This attack was first reported in 1988 by Norm Hardy~\cite{hardy1988confused}. This attack can happen when an application provides a public interface and access some sensitive resources. Other applications could use that interface to access the sensitive resources. The application providing access to the sensitive resource is called a confused deputy.
\par In 2011, the work in ~\cite{felt2011permission} mapped confused deputy attack with permission re-delegation attack. In particular, a careless developer may expose permission-protected resources through exported component. Other applications can access those resources through ICC to that component. \par In 2011, the first documented example of intentional permission re-delegation was presented by~\cite{schlegel2011soundcomber}. They developed Soundcomber, a Trojan with few and innocuous permissions, that can extract a small amount of targeted private information from the audio sensor of the phone and conveys information remotely without direct network access. This is the example of intentional permission re-delegation and illustrated in Figure \ref{soundcomber}.
\begin{figure}
	\centering
	\includegraphics[width=8cm, height=5cm, keepaspectratio]{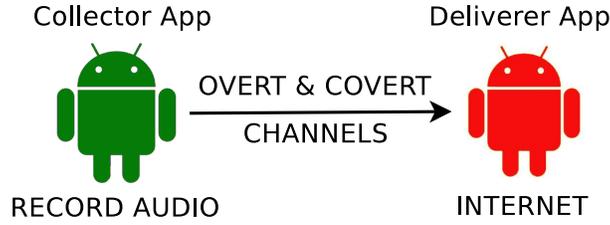}
	\caption{Soundcomber working architecture}
	\label{soundcomber}
\end{figure}
\par The Soundcomber example shows the difference between app collusion and confused deputy attacks. In app collusion the exposure of the  sensitive resource is intentional~\cite{heuser2016droidauditor}.
\subsection{Definition}
	Assume, $A$ be a set of all Android apps and $P$ be a set of all possible dangerous permissions in Android. \vspace{2mm}\newline
	Let, $a$ and $b$ are two apps with permission set $p_a$ and $p_b$ respectively such that:\newline
	\begin{center}
		$a,b \in A $ and $p_a , p_b \subset P$,
	\end{center}				
	Suppose, $a$ performs sensitive resource access that requires permission $\alpha$ to generate sensitive information $ \iota $ such that,
	\begin{center}
		$\alpha \in (p_a - p_b)$,
	\end{center}
	If $f(\iota)$ (perform any operation(s) on $\iota$) flows to $b$ through any number of apps, and $b$ performs sensitive resource access that requires permission $\beta$ to leak $ f(\iota) $ such that,
	\begin{center}
		$\beta \in p_b$,
	\end{center}
	\begin{center}
		\textbf{Then we say that $a$ and $b$ are colluding apps.}
	\end{center}
\subsection{Scenarios}
For the sake of completeness, we illustrate the collusion definition by three different scenarios based on app properties: 1) Among colluding apps, all are signed by the same signature; 2) Among colluding apps, all are signed with different signatures; 3) dynamic colluding apps.

\subsubsection{Among colluding apps, all are signed by same signature}
Android requires that all apps have to be signed with developer's certificate before they can be installed. Android uses this certificate to identify the developer of an app~\cite{signing}. 
There are some signing considerations as:\\
\begin{itemize}
\item App Modularity: Android allows apps signed by the same certificate to run in the same address space, if applications choose this, the system treats them as a single application. 
\item Code/Data Sharing: Android provides signature-based permissions enforcement, so that an app can expose functionality to another app that is signed with a specified certificate.
\end{itemize}
\par In this scenario, apps request a minimal permission set at the time of installation; and they can expand it later with the permissions of other apps with same signature. The shared permission set exposes sensitive information leakage path and thus emanate collusion.
\subsubsection{Among colluding apps, all are signed with different signatures}
Android applications signed by different signatures always run in different user address space. Such apps are not allowed to access each others data but can communicate with each other. Through this communication they can pass data. The general mode of such communication are Intents. Application do not need any specific permission to send Intent. In this scenario, collusion emanate if there exist some mode of communication between apps and they are sharing some sensitive information through it that is creating leakage path.
\subsubsection{Dynamic colluding apps}
This is a new type of threat called threat of split-personality behavior~\cite{maier2015game} where the attackers divide malware samples into a benign and malicious part, such that the malicious part is hidden from analysis by packing, encrypting or outsourcing of the code. In such cases the app appears as benign during analysis phase but while running on real device it become malicious. Similar case exists in case of app collusion, depending on the availability of an analysis system, malware can either behave benignly or load malicious code at runtime. In this scenario, static part of the app collude with dynamic part of the code to leak information. The dynamic part may be another app that get downloaded and installed using social engineering attack or it can be some dynamic code loading. The proof of concept of this type of collusion is demonstrated in~\cite{maier2014divide}. They modified the well known open source malware named AndroRat (Android Remote Administration Tool). This app connects to server and allow remote control of the device. The authors divide the app into two and developed AndroRAT-Split. They put the main service class into one app and added the activity class into another app to create collusion attack. Detection rate of original sample on VirusTotal is \textit{15/56} i.e. \textit{15} out of \textit{56} antivirus engines says that the app is malicious; whereas when AndroRAT-Split is analyzed  on Virus-Total, the detection rate is \textit{0/56} i.e. it is considered as benign app.	

\subsection{Challenges in Collusion Detection}
\label{challenges}
The detection of Android ICC based collusion faces many challenges. In the following, we summarize the major ones: 
\begin{itemize}
	\item How to characterize the context associated with communication channels with fine granularity?
	\item How to provide scalable solutions with minimum complexity to vet a large number of apps for possible collusion?
	\item How to define security policies for classification that reduce the number of false alerts?
	\item It's pretty hard to have a policy that is at the same time consistent and still provides realistic results without over-tainting.
\end{itemize}
The solutions to detect collusion in apps must be capable of doing analysis of multiple apps simultaneously aka inter-app analysis.
\section{Inter-Application Analysis}
This section discusses the main defense techniques for inter-app analysis: static, dynamic and policy based. 
\label{interAppAnalysis}
\subsection{Static Inter-App Analysis}
\label{staticInterAppAnalysis}
Static inter-app analysis consists of examining and auditing the code without executing it~\cite{tan2015securing}. Android apps are analyzed without really running them by inspecting the source code. Static analysis techniques act as a potential weapon for conducting the behavioral analysis of an application i.e. detecting whether an application is benign or malicious. It extensively explores data flows in a program and subsequently detect paths through which information can be leaked. It can be used to detect problems such as cross site scripting (XSS)~\cite{bhavani2013cross}, SQL injection~\cite{clarke2009sql}, buffer overflows~\cite{ceara2009detecting}, access control problems and many more. Resources and techniques of static inter-app analysis are detailed below.
\subsubsection{Resources}
\label{Resources}
The resources of Android apps from which information can be extracted in static analysis includes manifest file, dalvik byte code, libraries, etc~\cite{Schmeelk:2015:AMS:2746266.2746271}. \\The information obtained from manifest file includes the name of the package, list of components, list of permissions, version, etc. It also reveals information about intents and intent-filters used for communication. Level of API and libraries required by an application for execution is also mentioned in the file~\cite{manifest}. \\
Android apps are written in java and compiled to byte code. This byte code is further translated to dalvik byte code and stored in classes.dex (dalvik executable) file. This file reveals information about the structure of an application and methods used by it. It is analyzed to detect potentially malicious actions such as sending SMS to premium numbers, use of reflection or encryption, access of sensitive resources, etc.~\cite{karlsen2012study}. \\Java libraries can be statically analyzed in order to obtain data flow summaries of an application. It can be useful to determine malicious flow in an application.
 
\subsubsection{Techniques}
\label{Techniques}
The technique employed to perform static analysis depends on the depth and purpose of analysis. Various static analysis techniques used by researchers include taint analysis~\cite{scholz2008user}, dataflow analysis~\cite{reps1995precise}, entry point analysis~\cite{yang2012leakminer} etc. Some of the most prevalently used techniques are explained below.

Taint analysis is also known as user-input dependency checking~\cite{scholz2008user}. The concept behind taint analysis is that any variable altered by the user becomes tainted and is considered vulnerable. The taint may flow from variable to variable during a course of operations and if the tainted variable is utilized to perform some harmful operation, it becomes a breach in security. Taint analysis detects the set of instructions that are affected by user inputs. It helps in identifying sensitive information leakage.\\

Data flow analysis determines the information flow between various components. It is the essential analysis need to detect leakage of sensitive data. For instance, for a variable, it can detect all the possible sources of a variable's value i.e. where do values assigned to a variable come from, all the possible values a variable can possess, all the sinks where its value passes further, etc~\cite{dataflow}. Data-flow analysis can be of various types depending on the context of analysis ~\cite{Schmeelk:2015:AMS:2746266.2746271}. 
\begin{itemize}
	\item Context sensitive data flow analysis~\cite{arzt2014flowdroid} is an inter-procedural analysis technique. It examines target of a function call by focusing on calling context.
	\item Path sensitive data flow analysis~\cite{das2002esp} takes into account the branching statements. It analyzes the information obtained by the state obtained at conditional instructions.
	\item Flow sensitive data flow analysis~\cite{callahan1988program} considers the order of instructions in a program.
	\item Inter-procedural data flow analysis~\cite{myers1981precise} takes into account the flow of information between procedures. It is achieved by constructing call graphs
	\item Intra-procedural data flow analysis~\cite{sharir1978two} involves the flow of information within a procedure.
\end{itemize}

Entry point analysis~\cite{yang2012leakminer} helps in determining where a program starts its execution. It is very difficult to identify starting point due to the use of callbacks and multiple entry points. \\
Accessibility analysis~\cite{spyridi1990accessibility} contributes in evaluating the likelihood of following a path betwe~\cite{tan2015securing}en two components. It helps in building reachability graphs showing the path followed through methods for execution of an application. \\
Side-Effect analysis is performed to compute which variables of a method are affected by its execution~\cite{Schmeelk:2015:AMS:2746266.2746271}.


\subsection{Dynamic Inter-App Analysis}
\label{dynamicInterAppAnalysis}
Dynamic inter-app analysis refers to the analysis of a program by executing it~\cite{tan2015securing}. Android apps are examined and reviewed by actually executing them on real devices and emulators. Since static analysis does not portray the complete picture of an application, for example network data stored in the memory heap during run time is not available before executing app, obfuscated strings are hard to recognize from decompiled codes etc, therefore dynamic analysis is important to identify malicious applications, information leakage, sensitive data flows and vulnerabilities present in applications~\cite{petsas2014rage}.
\subsubsection{Resources}
\label{Resources}
The resources of Android apps from which information is extracted in dynamic analysis includes application framework, information about native code libraries, kernel parameters, cpu parameters, memory parameters, information about dynamically loaded libraries, etc. ~\cite{droidbox}. \\
Application framework is a software library which imparts basic structure for developing applications. It provides information about components and processes currently running and invoked API calls. It keeps record of which disk location belong to which file, keystrokes done and all the network inputs and outputs of function call. \\
Native library is a library that includes native code. Native code refers to a code written in C or C++ and compiled to machine code. Since native libraries are linked at run time, dynamic analysis is the suitable approach to identify the data flows in an application~\cite{ravitch2014multi}.\\
Kernel is the core of Android operating system. It is responsible for the management of hardware interactions. It provides information about the interactions with system protected resources which are not directly accessible~\cite{backes2014android}. \\
Other parameters of cpu, memory etc. can be captured dynamically when the program is running. 
\subsubsection{Techniques}
\label{Techniques}
Various dynamic analysis techniques used by researchers that include System hooking~\cite{costamagna2016artdroid}, Taint analysis~\cite{enck2014taintdroid}, Instrumentation~\cite{gupta2006dynamic}, System call tracing~\cite{pathak2011fine}, Debugging~\cite{machado2013mzoltar}, Code emulation~\cite{yan2012droidscope} etc.\\
System hooking involves altering or amplifying the functionalities of applications or components of application, by anticipating function calls, events and  transmitted messages between the components~\cite{sun2014design}. It assists in conducting dynamic analysis by intercepting and modifying API calls made by the target app. It is used to capture data flows, construct event ordering, record the parameters of passed messages and store values of run-time variables~\cite{backes2014android}.\\
Dynamic taint analysis~\cite{sarwar2013effectiveness} starts by tainting the data that is initiated from untrusted sources, specifically user supplied inputs. Later, these tainted variables are stored and whenever any of these variables are used for carrying sensitive data they are tracked down to detect sensitive paths ~\cite{yan2012droidscope}. \\
The term instrumentation pertains to the capability of monitoring or evaluating the performance of product and interpreting errors~\cite{amalfitano2011gui}. Android apps are instrumented to monitor actions of specific components such as logging number of times a particular service is called, etc. 
It is achieved by injecting smali codes. Smali is an intermediate representation of dalvik byte code which is inserted to keep log of actions of specific components. \\
In order to perform system call tracing, a system call tracer is embedded into the system that logs the invoked interrupts or APIs as the program runs on the system~\cite{szor2005art}. \\ In code emulation, the malicious code is executed on virtual machines with replicated CPU and memory management system, rather than real processor~\cite{szor2005art}.
\subsection{Policy Enforcement Based Analysis}
\label{rulebased}
Policy (a.k.a rule) enforcement based techniques make use of certain set of policies(rules) that are considered as normal or benign. These policies can be represented either in the form of regular expressions or any new policy language. The access of apps to any policy protected resource is verified against the predefined policy-set~\cite{enck2008mitigating}. Verification can be done statically on the intermediate program code and can also be enforced at install-time or run-time. If the resource access adhered to the policy-set, it is considered benign. Any violation is referred as malicious behaviour~\cite{xu2012aurasium}. The challenge posed by this defence mechanism lies in identifying, defining and maintaining the policy-set. It should not be very strict that may generate false-positives but at the same time it should not be too liberal to generate more false-negatives~\cite{enck2011defending}. 
\subsubsection{Resources}
The resources of Android apps from which information can be extracted in policy based analysis techniques depend on the nature of the policy-set and where they are applied~\cite{conti2010crepe}. For eg., if the policy set considers permissions and their corresponding API call then, manifest, dex files and libraries are suffice to extract relevant information. However, when policies are enforced at install-time or run-time, hooking or instrumentation need to be done. In later case, relevant information can be extracted from system parameter like registers, CPU etc.
\subsubsection{Techniques}
In order to protect users' private/sensitive data in Android, various solutions have been proposed based on controlling the access of sensitive resources. The access control can be apply at various system abstraction layers viz. kernel-layer, middle-ware layer and application layer~\cite{flaskdroid}. The access controls are of various type viz. Mandatory Access Control (MAC), Discretionary Access Control (DAC), Role Based Access Control (RBAC), Context Based Access Control (CBAC) and Attribute Based Access Control (ABAC)~\cite{rulebased,ubale2014analysis}.
\par In \texttt{MAC}, whenever app wants to access policy protected resource, Android kernel will verify the access against predefined rule-set. The access is allowed only if it is authorized. These rule-set is not modified by app or user. In \texttt{DAC}, user can define an access control list (ACL) on specific resources. These resources can be accessed when the owner provides permission. \texttt{RBAC} is based on the roles of an individual user. The user is assigning to different positions, with permissions to use the resources. The user can access sensitive data based on their assigned role. Till Android 5.1.1, once privileges are granted to the applications, they cannot be revoked. However, in many cases whether the application get a privilege or not depends on the user context and therefore \texttt{CBAC} comes into existence in Android. It has the capability to give privileges with dynamically granted or revoked to applications. In \texttt{ABAC}, granting privileges to the users is based on attributes which combine with the policies. Authorization relies on a set of operations that is determined by evaluating the attributes associated with the subjects, objects, and requested services.
\section{State of the art approaches}
\label{stateOfArt}
In the following section, we present different experimental approaches for the detection of intra and inter-application communication vulnerabilities. 
\subsection{Static Analysis}
This section briefly explains about the approaches conducting static analysis.
\subsubsection{MR-Droid: A Scalable and Prioritized Analysis of Inter-App Communication Risks~\cite{MR-Droid}}
		\textbf{Objective:} The paper aims to detect inter-app communication threats specifically intent hijacking, intent spoofing and collusion. Authors proposed a MapReduce based framework to scale up compositional app analysis. They also prioritized the identified ICC risks, based on the communication context of apps. \\\\
		\textbf{Methodology:} The MapReduce based approach is divided into two broad steps. In the first step, MR-Droid identifies ICC nodes (both sources and sinks), ICC edges (intent based ICC communication channels) and group inter-app ICCs that belong to an app pair using MapReduce. In the second step, risk assessment module of MR-Droid assigns risk levels to the app pairs based on the semantics and contextual information of the identified ICC channels. The risk assessment module detects the presence of risk and assigns ranking to the detected risk. Prioritizing risks helps to reduce false alarms. To validate the approach, authors manually analyzed around $200$ apps. \\\\
		\textbf{Dataset Used}: The dataset consists of  $11,996$ apps from $24$ popular app categories belonging to Google Play Store and $8$ apps from DroidBench 3.0~\cite{droidbenchnew} inter-app communication category. \\ \\
		\textbf{Limitations}: The paper suffers from the following limitations:
		\begin{itemize}
			\item The proposed approach can handle intent based ICC communications only. Therefore, security risks posed by other inter-app channels like content providers, shared preferences etc. cannot be detected.
			\item Currently, the approach can detect privacy leakage across two apps. However, leakage involving more than two apps are missed.
		\end{itemize}
\subsubsection{Detecting Inter-App Information Leakage Paths~\cite{bhandari2017poster}}
		\textbf{Objective:} The paper presented a model-checking based approach for inter-app
		collusion detection. The authors presented compositional app analysis to identify set of conspiring apps involved in the collusion. They developed $8$ colluding apps and contributed to DroidBench 3.0~\cite{droidbenchnew}. \\\\
		\textbf{Methodology:} The proposed approach is divided into four broad steps. In the first step, it leverages DARE~\cite{octeau2012retargeting} and IC3~\cite{octeau2015composite} to extract information related to ICC sources, sinks and, intent based communication channels. In the second step, data-flow analysis has been carried out to map sensitive information provided by sensitive API call to the outgoing intent followed by storing all the extracted information in the database. In the third step, PROMELA model is generated for each app. In the fourth step, model checking is done using the generated models and collusion detection property (\texttt{[](STATUS==SAFE)}) mentioned in the linear temporal logic (LTL) form. If the compositional model of apps do not satisfy the property, then that set of apps are declared as colluding apps set.  \\\\
		\textbf{Dataset Used}: The dataset consists of self-developed $8$ apps and contributed to DroidBench 3.0~\cite{droidbenchnew} under inter-app communication category.\\ \\
		\textbf{Limitations}: The paper suffers from the following limitations:
		\begin{itemize}
			\item The proposed approach can handle intent based ICC communications only. Therefore, security risks posed by other inter-app channels like content providers, shared preferences etc. cannot be detected.
			\item Not addresses the issue of scalability.  
		\end{itemize}
\subsubsection{Towards Automated Android App Collusion Detection~\cite{asavoae2016towards}}
\textbf{Objective:} The paper mentioned that collusion can cause information theft, money theft or service misuse. They defined collusion between apps as some set of actions executed by the apps that can lead to a threat. They proposed two approaches to identify candidates for collusion. One is rule based approach developed in Prolog and other is statistical based approach. \\\\
\textbf{Methodology:} In rule based approach, some features are used to identify colluding apps. These features include permissions, communication channels, and set of some actions viz. accessing sensitive information, sending information etc. The statistical approach consists defining probabilistic model, training of the model that means estimating the model parameter on the training set and validating the model on test dataset. Additionally the paper also presented that model-checking is the feasible approach to detect collusion in Android apps.\\\\
\textbf{Dataset Used}: The dataset consists of $\sim 9000$ malicious and $\sim 9000$ benign apps developed by Intel Security.\\ \\
\textbf{Limitations}: The paper suffers from the following limitations:
\begin{itemize}
	\item The rule based approach can be easily evaded using some evasion techniques like reflection, obfuscation etc.
	\item The statistical approach performance could be due to a bias of validation dataset towards the methodology.
	\item Not addresses the issue of scalability.  
\end{itemize}
\subsubsection{User-Intention Based Program Analysis for Android Security~\cite{elish2012user}} 
\textbf{Objective:} The paper proposed a data structure called ICC Map that statically captures cross-app information flow and based on self-made security policies classify communication among apps as collusion or no collusion. \\ \\
\textbf{Methodology:} ICC Map is a hash map data structure. It stores ICC entry and exit points that can be extracted  by scanning bytecode of source and target apps respectively. It is used to statically characterize the inter-app ICC channels among the Android apps. The detailed description of ICC Map is as follows:
\begin{itemize}
	\item ICC exit points refer to all Intent based ICC APIs like \texttt{startActivity(Intent i), startService(Intent i)} etc., user triggers like \texttt{onClick()} and APIs that retrieve private data such as \texttt{getAccounts(), getPassword()} etc.
	
	\item After extracting exit points, data dependence graph (DDG) of intra- and inter-procedural dependencies has been constructed.
	
	\item Then, from each ICC exit point, backward depth-first traversal on DDG is done to check if it involves private data or user trigger and store this information in \textit{SourceAppICCExitHashMap}.
	
	\item Then construct the data dependence graph (DDG) for the target app and perform forward	depth-first traversal on its DDG from each ICC entry point to find any critical operations	and store this information in \textit{TargetAppICCEntryHashMap}.
\end{itemize}
In SourceAppICCExitHashMap, each entry consists of source component name as key, and a list of ICC exit, sensitive data and user trigger as value. For example, \texttt{(compX,\{startService(Intent i), getDeviceID(), onClick()\})} represents one entry in the SourceAppICCExitHashMap, where compX is the component name that initiates the inter-app ICC call startService(Intent i) with sensitive data device ID included as part of the Intent of this call and onClick() as the user event to trigger this call. \\
Similarly, in TargetAppEntryHashMap, each entry consists of target component name as key, and a list of ICC entry, component protection and critical operation as value. For example, \texttt{(compY, \{onStart(), No, java.io.FileOutputStream.write(...)\})} represents one entry in the TargetAppICCEntryHashMap, where compY is the component name that receives the inter-app ICC call, and onStart() is the entry point of compY which is not protected (No) and has critical operation java.io.FileOutputStream.write(...). 
\par Given the above two hash maps, they connect inter-app ICC calls as follows:
\begin{itemize}
	\item First search for the source component name in SourceAppICCExitHashMap. The search results return its value(ICCExitName, SensitiveData, UserTrigger).
	\item Then, search the same with the target component in TargetAppICCEntryHashMap to	get its value(ICCEntryName, CompProtection, CriticalOperations).
	\item After that, connect the ICC exit point in the source component with its corresponding ICC entry	point in the target component. 
\end{itemize}
These operations provide the complete path of the ICC calls from the source to the destination across multiple apps. Authors call these paths as ICC links and entire data structure as ICC Map. ICC Map cannot be used for apps collusion detection but it helps to identify pair or group of communicating apps.
\par After this they define four rules/policies that are as follows:\\
Suppose component $ C1 $ in app $ P1 $ calls component $ C2 $ in app $ P2 $ , i.e., $ C1 \rightarrow C2 $ 
\begin{itemize}
	\item If the ICC exit point in $ C1 $ does not have a valid user trigger and the target component $ C2 $	is not protected by permission checking and has critical operation, then this ICC channel is classified as a high risk inter-app ICC channel. 
	\item If the ICC exit point in $ C1 $ has a valid user trigger and the target component $ C2 $ is not protected by permission checking and has critical operation, then this ICC channel is classified as a medium risk inter-app ICC channel.
	\item If the ICC exit point in $ C1 $ does not have a valid user trigger and the target component $ C2 $	is  protected by permission checking and has critical operation, then this ICC channel is classified as a medium risk inter-app ICC channel. 
	\item If the ICC exit point in $ C1 $ has a valid user trigger and the target component $ C2 $ is protected by	permission checking and has critical operation, then this ICC channel is classified as a low (or no) risk inter-app ICC channel.
\end{itemize}
Authors have also extended these rules to more fine-grained \textit{16} rules that include inter-app ICC call. The detail description of these \textit{16} rules are available in~\cite{elish2015user}.\\
Thus, based on ICC Map and a set of security policies they can differentiate between benign communicating apps and colluding ones. In their experiments, the proposed method can correctly detect \textit{97.9\%} of the \textit{1,433} malware samples. The false negative rate is \textit{2.1\%}, i.e., \textit{31} malware apps are misclassified as benign. \\\\
\textbf{Dataset Used}: \textit{1,433} malware apps collected from ~\cite{oberheide2012dissecting} and Virus Share. \textit{2,684} apps from Google Play market.\\\\
\textbf{Limitations:} ICC Map approach suffers from following limitations:
\begin{itemize}
	\item ICC Map cannot detect collusion through indirect communication channel such as shared files.
	\item ICC Map cannot capture the scenario that involves complex string operation like both apps read/write to files, however, the filenames are dynamically	generated using string operations. Collusion occurs through such scenario is missed by the proposed approach.
	\item The approach statically identifies the predicted risk level associated with the inter-app ICC calls, but it does not confirm the existence of the collusion.
	\item The proposed approach has difficulty in performing the analysis on programs that employ obfuscation techniques, dynamic code loading, or use of reflection.
\end{itemize}
\subsubsection{IccTA~\cite{li2015iccta}}
\textbf{Objective:} IccTA is a static taint analyzer to detect privacy leaks between components in Android apps. It claims to improve its precision of analysis by propagating context-aware information. \\ \\
\textbf{Methodology:} IccTA tool takes APK file (dalvik bytecode) as input and convert it into Jimple (soot's intermediate representation~\cite{lam2011soot}). After that, it extracts ICC links and related information like ICC call parameters, Intent filter etc. using Epicc~\cite{octeau2013effective} and also parses URIs (eg. scheme, host) to support Content Provider related ICC methods (eg. query) using IC3~\cite{octeau2015composite}. To extract ICC links, IccTA have to identify source and target components. Source components are the components that initiate ICC method and  target components are resolved by analyzing the values of Intent filter from \texttt{AndroidManifest} file of the app. It also needs to analyze bytecode because Broadcast Receivers may be declared at runtime. Then it stores all the extracted information into a database. Based on the extracted ICC links, IccTA modifies Jimple representation to directly connect the components to enable data-flow analysis between them.
\par IccTA handles three types of methods, ICC methods are replaced by an instantiation of the target component with the appropriate Intent. For callback methods, the tool takes care of both UI triggered event as well as callbacks triggered by Java or the Android system. To handle lifecycle methods, the tool generates a dummyMain method for each component in which it models, the entire lifecycle model of the component. IccTA leverages FlowDroid~\cite{arzt2014flowdroid} to build a complete control flow graph of the Android app under analysis. This graph allows to analyze the context (eg. the value of Intent) between two components. In the end, IccTA also stores the reported tainted path (leaks) into the database which can be reused in later analysis. IccTA achieved \textit{96.6\%} precision while analyzing privacy leaks from the samples of DroidBench and ICC-Bench. IccTA can perform inter-app analysis when used with APKCombiner~\cite{li2015apkcombiner} which is a static tool that scales down inter-app communication analysis to intra-app communication analysis. APKCombiner disassembled every app to obtain manifest and smali files using android apktool~\cite{apktool}, a reverse engineering tool. After that all files corresponding to different apps are combined together into a single directory and conflicts are resolved. \\ \\
\textbf{Dataset Used}: \textit{22} apps from DroidBench, \textit{15000} apps from Google Play Store, \textit{1260} apps from Genome Malware, \textit{16} apps from ICC-Bench.\\\\
\textbf{Limitations:} IccTA suffers from the following limitations:
\begin{itemize}
	\item IccTA resolves reflective calls only if their argument are string constants, which is not always the case.
	\item It cannot detect leak through multi-threading. It assumes the execution of threads in arbitrary but sequential order.
	\item It can miss leaks through native calls that their rules model incorrectly.
	\item It cannot handle rarely used ICC methods like \texttt{startActivities} and \texttt{sendOrderedBroadcastAsUser}.
	\item It cannot resolve complicated string operations which are generated using StringBuilder.
	\item The string analysis done by IccTA is within a single methods which may cause false alarms.
	\item IccTA cannot analyze apps of big size as it requires too much memory consumptions and system often gets hang.
\end{itemize}
\subsubsection{Automatic Detection of Inter-Application Permission Leaks in Android Applications: PermissionFlow~\cite{SbirleaPermissionFlow15}}
\textbf{Objective:} PermissionFlow is a single-app static analysis approach that handles attacks related to obtaining unauthorized access to permission-protected information. It focuses on three types of attacks viz. permission collusion, confused deputy and Intent spoofing. PermissionFlow uses taint analysis to capture the flow of permissions. \\ \\
\textbf{Methodology:} PermissionFlow consists of three major modules, i.e. Permission Mapper, Rule Generator and Decision Maker. The approach consists of identifying APIs whose execution leads to permission-checking. This is done through permission mapper. Then another module, rule generator will define rules for tainting. It considers the APIs selected by permission mapper, to be the sources of taint and define rules to capture their corresponding sinks. Then based on the information extracted from APK file of an app through apktool and dex2jar, PermissionFlow leverages another open-source tool named Andromeda to identify flows and components. Decision maker will allow or disallow the flow based on permissions.\\ \\
\textbf{Dataset Used}: PermissionFlow tests \textit{313} popular Android Market applications, and then identifies that \textit{56\%} of them use inter-component information flows that may require permissions. \\ \\
\textbf{Limitations:} Permission flow approach suffers from the following limitations: 
\begin{itemize}
\item Permission flow does not handle native code permissions.
\item Permission flow records a large number of false positives due to the checking of redundant permissions and data dependent checks.
\item Permission flow gives false negatives for the apps that transfer protected information between components before returning it. This is due to the use of Implicit intents as it prevents identification of the class names for invoked \texttt{Activity}.
\end{itemize}
\subsubsection{FUSE~\cite{ravitch2014multi}}
\textbf{Objective:} FUSE proposed a approach that starts by single-app static analysis accompanied with lint tool followed by multi-app information flow analysis. Lint tool is used to mitigate limitations of static analysis. They demonstrated limitations of single-app analysis by detecting more information flow paths in multi-app analysis. \\ \\
\textbf{Methodology:} FUSE works in two broad steps. Single-app analysis, followed by multi-app analysis based on violation of specified security policies. The first step takes an appkit (collection of apps) as input, analyze each app individually to produce extended manifest data structure. In the second step, all the extended manifests are combined and collusion is checked based on violation of specified security policies. The detailing of all the steps are as follow:
\par Single-app analysis: In this step, each application in the appkit is individually analyzed to create its corresponding extended manifest data-structure. The extended manifest represents the internal information flow graph from application inputs (sources) to application outputs (sinks). According to FUSE, sources are the inputs to each component or permission protected resources. Sinks are the means by which a component can send (possibly sensitive) information to another component or to the outside world. In FUSE, a version of Andersen's analysis~\cite{andersen1994program} is used to compute a call graph and determine reachable methods in each application. The analysis also considers Android Framework and Java libraries along with the application. The application is tainted with taint labels at every source and if these labels reach to sinks, data leakage path is flagged.
\par Multi-app analysis: In this step, FUSE takes the entire appkit along with all the extended manifests as input. The output is the multi-app graph with the flow of information between the applications. In the graph, set of all the permissions and each sources and sinks present in any application becomes the node. There must be an edge from source or permission to the sink, whenever there is the flow of information between them. There also exist the edges from sink to component, if sink can send IPC message to that component.
\par FUSE defines coarse-grained information flow assertions based on permissions combination. The multi-app graph is checked against these assertions. User is alerted if there is any violation of the assertion occurs in the graph. FUSE also uses security linter tool, to overcome the limitations of static analysis. The tool issues warning if there exists problem like, component hijacking, dynamic registration of broadcast receiver, use of insecure credentials, presence of reflection, unused permissions or writing to public files.\\\\
\textbf{Dataset Used}: \textit{189} applications drawn from Nexus 4 running Android 4.4.2. \textit{1124} applications of F-Droid updated till May,2014. \textit{1260} applications from Genome project dataset, applications from DroidBench. \\\\
\textbf{Limitations:} FUSE suffers from following limitations:
\begin{itemize}
	\item The biggest limitation of the approach is that it is not publicly available to test. It is designed for commercial purposes.
	\item Defining coarse-grained information flow assertions leads to many false alert as the behavior of the application is not considered.
	\item It does not support all versions Android APIs, 
\end{itemize}
\subsubsection{Amandroid \cite{wei2014amandroid}}
\textbf{Objective:} Amandroid is a static analysis tool, that has the capability of calculating all objects' points-to information in a both flow and context-sensitive way. 
This tool detects whether there is any information leakage from a sensitive source to a critical sink; by providing an abstraction of the app's behavior. 

\textbf{Methodology:} Amandroid proceeds by converting an app's Dalvik bytecode to an intermediate representation (IR) for subsequent analysis. Then, Amandroid  generates an environment model that emulates the interactions of the Android System with the app to limit the scope of the analysis for scalability. Amandroid builds an inter-component data flow graph  (IDFG) of the whole app. IDFG  includes the control flow graph; that tracks the set of object creation sites that reach each program point.  The core component is to  build a precise IDFG  of the app; the flow-sensitive and context-sensitive data flow analysis to calculate objects points-to information is done at the same time with building inter-procedural control flow graph.  Amandroid builds the data dependence graph on top of the IDFG, then it induces explicit information flow. 
This framework provides an abstraction of the app's behavior, and can be used for a number of useful security analysis as data leak detection, data injection detection, and detection misuse of an API.

\textbf{Dataset used:} Amandroid is tested on $753$ Google Play apps by the Epicc group, and $100$ potentially malicious apps from Arbor Networks.

\textbf{Limitations:} Amandroid has the following limitations: 
\begin{itemize}
	\item Amandroid has limited capability to handle exceptions. Amandroid may not detect an exception, when an app has a security issue where the core of an execption handler plays a role. 
	\item Amandroid does not handle concurrency and reflections. An app may have multiple components and then may run concurrently; and when multiple components interleave this may induce some security issues. 
\end{itemize}
\subsubsection{Android Taint Flow Analysis for App Sets~\cite{Klieber:2014:ATF:2614628.2614633}}
\textbf{Objective:} DidFail conducts static taint analysis of Android apps by augmenting FlowDroid and Epicc tools to detect intra-component and inter-component information flow in a set of apps. It performs analysis in two phases where the first phase determines information flow within the app and second phase determines flow across the apps.\\
\textbf{Methodology: }DidFail accepts a set of apps as the input. The analysis is performed in two phases:
\begin{itemize} 
	\item The first phase constitutes of 4 steps - TransformAPK, FlowDroid (modified), Dare and Epicc. 
	\begin{enumerate}
		\item \textit{TransformAPK}: In this step, each APK is modified by using Soot. Initially, APK is converted to an intermediate representation known as \texttt{jimple}. Later, all the send intent method calls are located and just before the method call, a new method call is inserted that provides a unique ID to the sent intent. The jimple code is then repackaged into an APK and passed as an input to the next step.
		\item \textit{Dare}: This tool accepts transformed APK as an input and produces retargeted java class files as an output. 
		\item \textit{Epicc}: This tool accepts retargeted java class files and transformed APK as an input and provides parameters of sent and received intents such as \texttt{action, category} as an output to the second phase.  
		\item \textit{Modified FlowDroid}: DidFail has modified FlowDroid by adding few intent method calls as sources (onActivityResult()) and sinks (setResult()). It has also added code to analyze \texttt{putExtra} call for the intents that are uniquely identified by ID in TransformAPK step. FlowDroid accepts transformed APK as the input and conducts taint analysis. It provides flows within the components of an app as the output.
	\end{enumerate}
	\item The first phase identifies intents as tuples. For example, $I<C_1,C_2,ID>$ where $C_1$ = component that sends the intent, $C_2$ = component that receives the intent and ID = unique identifier of the intent. This phase identifies flows within an app and pass these flows as an input to the second phase.
	\item In the second phase, inter-app communication among the set of apps is resolved i.e. an intent sent by an app is matched with the intent-filters of other apps to identify the receiver. Once the receiver is identified a flow from source$\rightarrow$sink is detected.
\end{itemize} 
\textbf{Dataset Used: }DidFail is tested on two app sets where the first app set contains three apps developed by the authors and second app set contains three apps taken from Droidbenchmark.\\
\textbf{Limitations: }DidFail suffers from the following limitations:
\begin{itemize}
	\item DidFail does not handle native calls and reflection.
	\item DidFail focuses only on \texttt{Activity} component of Android app. It does not handle service, broadcast receiver and content provider.
	\item DidFail cannot detect flow of information when static fields are used as a source or sink for intents i.e. it misses the flow if an intent reads information from static field. 
	\item If the tainted information propagates through a chain of apps, then DidFail fails to detect the flow. 
\end{itemize}
\subsubsection{Analyzing Inter-Application Communication in Android: ComDroid~\cite{chin2011analyzing}}
\textbf{Objective:} ComDroid is a tool that detects application communication vulnerabilities and could be used by developers and reviewers to analyze their own applications before release. The main purpose of this tool comes from the fact that Android's message passing system can become an attack  if used incorrectly (personal data loss, information leakage, phishing, etc.) These vulnerabilities stem mainly from the fact that Intents can be used for both intra and inter application communication. \\\\
\textbf{Methodology:} ComDroid considers two types of analysis: Intent analysis and Component analysis.
\par In Intent analysis, ComDroid statically analyzes method invocation to a depth of one method call. In this way it performs flow sensitive intra-procedural static analysis, with a limited inter-procedural analysis. This tool parses dalvik files and tracks the state of intents, registers, sinks, intent-filters, and components.  For each method that uses intents, this tool can track the value of each string, class, intent and intent-filter. For each Intent object, ComDroid tracks the following:
\begin{itemize}
	\item whether the intent has been made explicit;
	\item whether the intent has an action;
	\item whether the intent  has any flags set; and
	\item whether the intent  has any extra data.
\end{itemize}
When it detects that an implicit intent being sent with weak or no permission requirements, ComDroid issues a warning as this situation is eavesdropping prone. There are two types of warnings viz. with data, and without data, in order to distinguish action based attacks from eavesdropping. 
\par In Component analysis, ComDroid examines application's manifest file to get components and translates dalvik instructions to get information about each component. ComDroid treats activities and their aliases as separate components because an alias field can increase the exposure surface of the component. 
It generates a warning about a potential intent spoofing attack, when it detects that a public component is protected with no permission or a weak permission. ComDroid also issues warnings for receivers that are registered to receive system broadcast actions (that are actions sent by the system).
\par In order to resolve these warnings, a solution proposed by authors was to add a call to \texttt{android.content.Intent.getAction()} to verify that the protected action is in the Intent (authentication of the sender of the Intent). This differs from other Intent spoofing attacks where the solution is to make the component private. \\\\
\textbf{Limitations}: ComDroid suffers from the following limitations:
\begin{itemize}
	\item False Negatives: ComDroid tracks Intent control flow across functions, and did not distinguish between paths through if and switch statements.  For instance, an application might make an Intent implicit in one branch and explicit in another, ComDroid would always identify it as explicit. 
	\item  Privilege Delegation: ComDroid does not detect privilege delegation through pending Intents and Intents that carry URI read/write permissions.  
	\item  Verification of the existence of attacks: ComDroid issues warnings and not verify the existence of attacks.  For instance, some components are intentionally made public for the purpose of inter-application collaboration. It is not possible to infer the developer's intention  when making a component public. It is the role of the developer to verify the veracity of the warnings. 
\end{itemize}
\subsection{Dynamic Analysis}
This section briefly explains the proposed tools and approaches that conducts analysis dynamically.
\subsubsection{IntelliDroid~\cite{wong2016intellidroid}}
\textbf{Objective:} IntelliDroid is a generic tool that generates input specific for a dynamic analysis tool to perform analysis more precisely by reducing false positives. Instead of static or dynamic analysis, this work proposes targeted analysis. It is achieved by preliminarily doing background study about the dynamic analysis tool and static analysis of the application given as input to the dynamic analysis tool. It helps in triggering target APIs and consecutively leads to more efficient and effective dynamic analysis.\\\\
\textbf{Methodology:} Android apps contain multiple event handlers, which when triggered in a particular sequence with specific inputs, reveal malicious behavior. This environment and input is provided by IntelliDroid to dynamic analysis tools. IntelliDroid acts in \textit{6} steps viz. Specifying target APIs, Identifying paths to target APIs, Extracting call path constraints, Extracting event chains, Determining run-time constraints and Input-injection to trigger call paths.
\par In the first step, APIs to be targeted are identified by analyzing either API methods~\cite{enck2014taintdroid}, system calls~\cite{tam2015copperdroid} or low-level events~\cite{shabtai2012andromaly}.
\par In the second step, paths to the targeted APIs are discovered by conducting static analysis. IntelliDroid obtains information about components of an application and its lifecycle methods by reading its manifest file. It identifies entry-points of an application and create a partial call-graph to look for initialization of callback listeners. It adds circumvented listener methods to the entry-points list and creates a new call graph. This process is repeated recursively till no more entry points are found. By traversing path from event handler's entry point to target API invocation, target paths are extracted for every target API.
\par In the third step, call path constraints are determined by conducting control and data-flow analysis on the control flow graph (CFG) in forward direction. If more than one path exists from one-method invocation to other in CFG, IntelliDroid combines the constraints of each path by using logical OR operator. For the cases where extracted constraints are return values of other method invocation, they are added with main path constraints by using logical AND operator.
\par The reason for executing fourth step (extracting event chains) is that the path constraints can be heap variables whose value cannot be determined statically by observing entry-points. To determine their values, the lines in the code containing the heap variable definition are looked for. The event handler containing the heap variable definition are tracked and stored. The route from event handler to heap variable store statement becomes supporting path and its constraints becomes supporting constraints. Their value is determined at the time of resolving constraints ans subsequently used for storing path constraints.
\par In the fifth phase i.e. determining run-time constraints, the value of variables that are still unresolved are obtained at run-time, just before the event injection.
\par In the sixth step, finally the input fulfilling constraints are injected to trigger the call paths. The component of IntelliDroid responsible for injecting input consists of a computer attached to a device. Communication between them occurs through IntelliDroidService. The static part of IntelliDroid is responsible for specifying inputs for the targeted APIs. It supplies these inputs to dynamic part which is accountable for inserting the inputs at device-framework interface of Android.
\par For the static analysis, source code is not used. The APK files are unpacked using Dare~\cite{octeau2012retargeting} and APKParser~\cite{apkparser}. The Java bytecode is then passed to static part which uses WALA static analysis libraries~\cite{wala}. For dynamic analysis, Z3 constraint solver~\cite{de2008z3} is used. The IntelliDroidService client program is executed using Python.
\par IntelliDroid is tested with TaintDroid, a dynamic analysis tool. On an average \textit{72} inputs have been injected. Out of \textit{75} malware instances, IntelliDroid was successful in identifying \textit{70} instances. It is observed that \textit{138.4} seconds on an averages is required per application.\\\\
\textbf{Dataset Used}: It has been tested on \textit{1260} malware samples from Malware Genome Project~\cite{zhou2012dissecting} and \textit{1066} benign apps from Android Observatory~\cite{barrera2012understanding}.  \\\\
\textbf{Limitations:} IntelliDroid suffers from the following limitations:
\begin{itemize}
	\item IntelliDroid does not handle implicit Intents, Content Providers and native code.
	\item The extracted constraints are sometimes very complex such as trignometric functions. It cannot be resolved by constraint solver. Currently, human intervention is required to solve such constraints.
	\item IntelliDroid partially handles reflection as it cannot identify the path constraints after the reflected call.
	\item IntelliDroid is not capable of creating inputs for encrypted and hashed functions.
\end{itemize}
\subsubsection{IntentDroid~\cite{hay2015dynamic}}
\textbf{Objective:} IntentDroid is a framework that dynamically examines Android apps for IAC (Inter Application Communication) related integrity vulnerabilities such as custom uri's, payloads in IAC messages etc. It created attack scenario for \textit{8} vulnerabilities viz. Cross-Site Scripting, SQL Injection, Unsafe Reflections, UI (User-Interface) Spoofing, Fragment Injection, Java Crashing, Native Memory Corruption and File Manipulation. It analyzes Activity component of apps by implementing attack scenarios in a way to obtain effective path coverage with minimum overhead.\\\\
\textbf{Methodology: }IntentDroid tests the applications in three phases viz. Instrumentation, Testing and Reporting. 
\par In the instrumentation phase, the app under analysis is instrumented to store library calls and access to user-supplied data. The app is reverse engineered through apktool to extract its manifest file. Manifest file is parsed to extract public components. IntentDroid specifies three cases to call any activity as public:
\begin{itemize}

		\item if the activity is exported via Intent filter(s);
		\item if the activity access does not require any permissions (system or signature); and
		\item if any unvalidated data which is originated from any public component, passes through it;

\end{itemize}
These activities communicate via Intents and therefore all Intents form a set of IAC input points for the Testing phase. 
\par In the testing phase, to detect whether a vulnerability exists in the app or not, IntentDroid has created attack scenarios. Testing occurs in three steps Monitoring, Testing and Exploration. During monitoring, IntentDroid sends a message to the app under test. It uses system-level hooks to records all the security concerned APIs called by the app and custom fields (such as \texttt{getStringExtra}, \texttt{getFieldExtra}, etc.) accessed by it. IntentDroid analyzes the app with direct and indirect access to custom fields. For direct access, IntentDroid iteratively detects the extra fields of the Intent and looks for the behavior of the app. If any extra field is observed, that IAC input point migrates to testing phase. For indirect access, bundle object created for the message sent by IntentDroid is observed by installing a monitor in \texttt{Intent.getBundle()} method. It checks the extra and data fields for payloads. If any inserted payload is found, bundle is considered relevant and exploited further by implementing attack scenarios. After monitoring the app undergoes testing. The app is tested for implementation of an attack scenario on an IAC input point by sending probe request. If the result is positive, that attack scenario is implemented. In the end during exploration, boolean variables are analyzed to detect the path followed by an Intent. It introduces two terminology for boolean variables viz. Independence and Dominance. A boolean variable is Independent, if its execution is not dependent on any other boolean variable. On the contrary, a boolean variable is Dominating, if its execution decides the application of other boolean variable (nested variables). Boolean variable analysis reflects the handling of incoming data by the Intent. 
\par In the Reporting phase, IntentDroid reports the number of vulnerabilities present in an app after implementing all the possible attack scenarios on the app. 
\par For evaluating IntentDroid, apps in the test dataset are manually tested by professional ethical hacker through brute-force fuzzing tool. It detected \textit{163} IAC vulnerabilities across \textit{80} apps. IntentDroid is able to detect \textit{150} IAC vulnerabilities giving a recall rate of \textit{92\%}.\\\\
\textbf{Dataset Used}: IntentDroid is tested on the dataset of \textit{80} apps, out of which \textit{4} are enterprise apps, \textit{4} are shipping apps and remaining \textit{73} are the most popular Google Play apps. These apps are tested on Samsung Nexus 5 device with Android 4.4 installed on it.\\\\
\textbf{Limitation: }IntentDroid suffers from the following limitations:
\begin{itemize}
	\item IntentDroid does not test Services, Broadcast Receivers and Content Providers for IAC vulnerabilities.
	\item IntentDroid does not consider multi-app attack.
\end{itemize}
\subsubsection{TaintDroid~\cite{enck2014taintdroid}}
\textbf{Objective: }TaintDroid is a security framework that dynamically detects sensitive information leakage in ICC between Android apps. TaintDroid extends the functionality of Android operating system to record the flow of confidential and vulnerable data through installed applications. \\\\
\textbf{Methodology: }It combines four fragments of taint dispersion viz. Variable Level, Message Level, Method Level and File Level. Variable level capturing is done for single app analysis. Variable taint tags are stored adjacently to variables in memory. Message level capturing is done for tracking communication between applications. One taint tag per message is stored which is the combination of variable taint tags. Method level capturing is done for native libraries granted by system. File level capturing is done to guarantee data preserves their taint markings. One taint tag per file is stored.
\par TaintDroid divides the Android architecture in three modules viz. Interpreted Code, Userspace and Kernel. It considers a scenario in which a message is transmitted from source app to destination app, where source app is assumed to be trusted and destination app is assumed to be untrusted. 
\par Interpreted code module of the source app taints (labels) the data originated from confidential and vulnerable sources (such as GPS coordinates) as taint sources in a trusted application. The assigned taint labels are stored in \textit{Virtual Taint Map} present in Userspace module. 
\par Userspace module includes Dalvik VM interpreter (which is invoked by native methods) and Binder IPC library. TaintDroid modifies Binder IPC library so that in case of an ICC, the parcel transmitted between two apps carries a taint tag which is a combination of taint markings of all the data carried inside the parcel. The customized Android platform tracks the flow of tainted data through dynamic taint tracking. In dynamic taint tracking, the labels are assigned transitively to components (such as IPC messages, variables etc.) when sensitive information propagates through them. When the tainted data of trusted app is to be sent as an ICC message, the data is first transferred to modified Binder IPC library. It creates a parcel and ensures that parcel holds a tainted tag, representing the combination of all the tainted data tags inside the parcel. The parcel is sent to untrusted application via Kernel. 
\par On the receiver side, modified Binder IPC library extracts the data from the parcel and sends it to Dalvik VM Interpreter. The flow of tainted data is monitored. Whenever tainted data leaves the system either by transmission over the network or by any tainted sink, the scenario is logged and reported to the user immediate. The logged information includes labels of data, receiver of the data and the application culpable for sending the data. 
\par TaintDroid reported that on an average two-third of the considered apps are leaking sensitive data. TaintDroid incurs \textit{14\%} of performance overhead on CPU-bound micro-benchmark.\\\\ 
\textbf{Dataset Used}: \textit{30} most popular android apps are selected from \textit{12} categories of Android Market. By applying TaintDroid, \textit{65} scenarios of information misuse across \textit{20} apps has been identified. Out of \textit{1130} logged TCP connections, \textit{105} has been found responsible for carrying tainted data out of the system. \textit{15} out of \textit{30} apps have been found leaking user's location to advertising servers. \textit{7} applications have been detected leaking user's device ID.\\\\
\textbf{Limitations: }TaintDroid suffers from the following limitations:
\begin{itemize}
	\item TaintDroid only handles data flows. It does not considers control flows.
	\item TaintDroid is not able to tag native code which leaves many sensitive sources untouched.
	\item In case of File level tracking, storing one taint tag per file gives a lot of false positives.
\end{itemize}
\subsection{Policy Enforcement Based Analysis}
{\color{blue}{
		\subsubsection{Collusive Data Leak and More: Large Scale Threat Analysis of Inter-app Communications\cite{bosu2017collusive}}
		\textbf{Objective:} The paper presents a tool named DIALDroid (Database powered ICC Analysis for Android). To the best of our knowledge, this is the first state-of-art that proposed large scale detection of collusion and privilege escalation. They also provide the first inter-app collusion real-apps benchmark of $30$ apps. Till now, this is the most efficient tool available in the literature for inter-app vulnerability detection~\cite{MR-Droid}. \\\\
		\textbf{Methodology:} In this paper, authors proposed DIALDroid that works in four broad steps. In the first step, permissions and intent-filter attributes are extracted from the manifest file and ICC entry/exit points are identified. In the second step, static taint analysis is performed to determine paths from sensitive sources to the intents being sent and intents received to sensitive sinks. It leverages Flowdroid\cite{arzt2014flowdroid} to conduct dataflow analysis at high precision. In the third step, all the extracted data is organized in mysql database comprising of 42 tables. The relational database provides scalable and efficient storage. To reduce the computational complexities, DIALDroid filters out ICC communications that are not sensitive. Finally, security policies are implemented using sql queries to detect the presence of collusion or privilege escalation. 
		\par They improved the preciseness of intent discovery by implementing incremental callback analysis. In dataflow analyzer, if any app takes more than $5$ minutes, DIALDroid resets the analysis by decreasing precision to maintain the trade-off between performance and precision. To avoid deadlocks, the app is analyzed for maximum $20$ minutes. The crash rate of DIALDroid is very less than IccTA+APKCombiner~\cite{li2015iccta} and it is more accurate than~\cite{li2015iccta,covert}.
		\\ \\
		\textbf{Dataset Used:} The dataset consists of 110,150 apps which includes 100,206 most popular Google play apps and 9,944 apps from Virushare. DIALDroid is also analyzed on Droidbench apps and ICC bench apps. \\\\
		\textbf{Limitations:} DIALDroid suffers from the following limitations:
		\begin{itemize}
			\item DIALDroid resolves reflective calls only if their argument are string constants, which is not always the case.
			\item DIALDroid analyzes an app only for 20 minutes. If an app takes more than that time, DIALDroid stops analysis.
			\item DIALDroid can handle intent based ICC communications only. Therefore, security risks posed by other inter-app channels like content providers, shared preferences etc. cannot be detected. 
\end{itemize} }}
\subsubsection{Intersection Automata based Model for Android Application Collusion~\cite{intersection}}
\textbf{Objective: }This is a static inter app analysis tool that can take multiple apps simultaneously for analysis and detect potentially colluding apps.\\\\
\textbf{Methodology: }In this paper, authors proposed a novel automaton framework that allows detection of intent based collusion among apps. The presence of collusion is detected by intersecting application and policy automata. Application automaton depicts intent-based communication among apps. Policy automaton have policies like if access of an API that requires \texttt{READ\_SMS} permission in one app is followed by an API call that requires \texttt{SEND\_SMS} permission in another app. The policy is searched in the application automaton and if the match is found, the tool will check the presence of \texttt{READ\_SMS} permission in the second app. If it is not found then the tool declares the presence of collusion otherwise no collusion. 
\par The detection framework operates at the component-level. They tested their approach on 21 apps by taking all possible
combinations (two at a time) and successfully detected presence/absence of collusion among them. Time and space complexity of the proposed tool is $O(n)$ where $n$ is the sum of all the components in applications under analysis. \\\\
\textbf{Dataset Used}: 3 applications from DroidBench inter-app communication category, self-developed 14 applications and 4 applications from Google Play Store. \\ \\
\textbf{Limitations: } The tool suffers from the following limitations:
\begin{itemize}
	\item The false alarm rate is very high as the tool is not performing any data-flow analysis. If there is intent communication between two apps without any data transferred, the tool raise warning of collusion.
	\item Only intents are considered as a means of communication. 
\end{itemize}
\subsubsection{Flexible and Fine-Grained Mandatory Access Control on Android	for Diverse Security and Privacy Policies ~\cite{flaskdroid}}
\textbf{Objective:} FlaskDroid is policy-driven tool that provides security for kernel resources (like files, IPC, etc.) as well as  middleware resources (like Intents, Content Providers, etc.). The security enforcement is through providing mandatory access control on both middleware and kernel layers of Android simultaneously. They extended Android's middleware layer with type enforcement and present a new policy language to capture the semantics of this layer.  \\ \\
\textbf{Methodology}: FlaskDroid plants various \texttt{Object Managers} at middleware and kernel layer that are responsible for assigning security context to objects. Related policies are managed by security servers deployed at different layers. The object manager makes access control decisions by using security servers at their respective layer. Also the deployed policies at both the layers are synchronized meaning change of policy in one layers, automatically reflect in another layer. Following are the major components of FlaskDroid:
\begin{itemize}
	\item SE Android Module: SE Android module restricts the privileges of root account to constrain the file-system privileges of the app. It is also responsible for restricting apps from bypassing middleware level policy enforcement check. For e.g., it restricts app from directly accessing the contacts database file instead, the app must access contacts via ContactsProvider app.
	\item Userspace Security Server: It is responsible for taking policy decisions for all userspace access control.
	\item Userspace Object Managers: In FlaskDroid, middleware services and apps act as Userspace Object Managers (USOMs) for their re-
	spective objects. Currently it comprises of 136 policy enforcement points.
	\item Context Providers: A context is the current security requirements of the device. It is derived from various criteria, such as physical, the state of apps and the system. Context Providers are the plugins to Userspace Security Server that allows control of contexts and their definitions. \\
\end{itemize}
\textbf{Dataset Used}: FlaskDroid is evaluated on the apps collected from Malware Genome~\cite{genome} and Contagio minidump~\cite{contagio}. The authors also developed synthetic apps that exhibits root exploit, over-privilege, information leakage, sensory malwares, confused deputy and collusion attacks. \\ \\
\textbf{Limitations: } FlaskDroid suffers from the following limitations:
\begin{itemize}
	\item Access control rules are human user trail based. Therefore, revision of rules is time-comsuming process. Also, limited human trials cannot guarantee full coverage of possible access control rules.
	\item Many false alarms while detecting confused deputy and collusion attacks. As FlaskDroid relies on application inputs/outputs and does not consider the information flow within apps.
	\item Simultaneous analysis of multipe apps is not provided by FlaskDroid.
\end{itemize}
\subsubsection{XManDroid: A New Android Evolution to Mitigate Privilege Escalation Attacks~\cite{bugiel2011xmandroid}}
\textbf{Objective:} XmanDroid (eXtended Monitoring on Android) is a dynamic framework that extends the monitoring mechanism of Android to detect and prevent application-level privilege escalation attacks. It is based on runtime system-centric policies. 
Two types of application-level privilege escalation attacks are handled by XmanDroid viz. Confused Deputy attacks and Colluding attacks.\\\\
\textbf{Dataset Used}: XManDroid developed their own dataset that consists of seven apps. These app set exhibit privilege escalation attack through ICC communication links and three types of covert channels viz. synchronized adjustment and reading of the voice volume, change of the screen state and change of the vibration settings. \\ \\
\textbf{Methodology}: XManDroid consists of three elements: 
\begin{itemize}
	\item Application Installer: It is responsible for installation and uninstallation of applications. It makes use of package manager for incorporating the changes and rebuild a new state, whenever any new application gets installed. 
	\item System Policy Installer: It is responsible for the installation of explicitly defined list of system policies in the Android middleware. 
	\item Runtime Monitor:  It is responsible for enforcing mandatory access control in Android like permissions are checked at this interface, take decisions whether to allow an ICC or not based on the information about installed apps and their communication. Whenever a request for an ICC call reaches, it is either approved or disapproved by reference monitor after validating it with policies database and whether the given ICC call leads to privacy leak or not.
\end{itemize}
\par System representation is done using graph schema where UID assigned by the system to an app is a vertex and the information about exchanged intents are the edges. By applying the rules of graph theory, transitive transfer of information is detected during ICC. This graph is used for defining rules in  system policies. \\ \\
\textbf{Dataset Used}: XManDroid developed their own dataset that consists of seven apps. These app set exhibit privilege escalation attack through ICC communication links and three types of covert channels viz. synchronized adjustment and reading of the voice volume, change of the screen state and change of the vibration settings. \\ \\
\textbf{Limitations: } XmanDroid suffers from the following limitations:
\begin{itemize}
	\item False Positives: XManDroid suffers from high positives if the app under analysis is over-privileged. The defined system-policies are not tuned properly.
	\item Attack at kernel level: XManDroid cannot handle privilege escalation attacks done at kernel level that can exploit the system to gain root access.
	\item Single app analysis is missing: XManDroid cannot detect malicious app as applications within a single sandbox have equal privileges and cannot perform privilege escalation.
\end{itemize}

\subsection{Case Studies}
This section presents various studies done to demonstrate the serious effects of information leakage in android apps. 
\subsubsection{Case of Collusion: A Study of the Interface Between Ad Libraries and their Apps~\cite{book2013case}}
\textbf{Objective: }In this study, API calls used by Ad Libraries to communicate with host applications are analyzed. Host applications have access to sensitive and private user data. API calls are capable of transmitting demographic data about user, which is of great interest for Ad agencies. Therefore, this interface presents a serious impact of user's private information leakage.  \\\\
\textbf{Methodology: }Initially, manual identification of \textit{103} individual and analytic libraries are done. Then, apps are disassembled using dedexer~\cite{dedexer} followed by app parsing, to detect all the API calls (using package name) that occurred between apps and any ad library. Later, all the captured API calls are recorded to keep a track of calls which were actually used from the considered dataset. Frequency of each call is also calculated and recorded. The group of detected API calls are assembled to recreate API of Ad libraries. Recreation is followed by detection of privacy related API calls through manually examining each API call using method name and parameters. The study found that user private data is leaked and stored in databases. These demographic data can be correlated to map a user to a real world person.\\\\
\textbf{Dataset Used}: Dataset consists of \textit{114,000} apps downloaded from Google Play Store. APIs of \textit{103} ad libraries used by apps in the dataset are reconstructed. Top \textit{20} ad libraries used in \textit{64000} applications have been analyzed to detect privacy leakage.\\\\
\textbf{Limitations: } The study suffers from following limitations:
\begin{itemize}
	\item The study only considers API calls. API calls are not the only source of communication between ad libraries and apps. Communication can also be performed through direct manipulation of class variables, shared memory etc. 
	\item In case of obfuscation, where method names are altered, this method is not capable of identifying privacy related API calls.
	\item The libraries which acts as a intermediate to transfer information between an app and its library are known as Ad mediation libraries. They are not considered in this study.
	\item It also omits small libraries which can also be a source of privacy leakage. 
\end{itemize} 
\subsubsection{Analysis of communication between colluding applications~\cite{marforio2012analysis}}
\textbf{Objective:} This paper focuses on evaluating two major channels used for collusion named covert and overt channels. This work aims to quantify the severity of app collusion attack threats by computing throughput, bit-error rate and expected synchronization for every channel.\\\\
\textbf{Methodology:} Covert and overt channels are implemented by executing them on Nexus one or Samsung Galaxy S smartphones. \textit{5} overt channels are implemented viz. Shared Preferences, Internal Storage, Broadcast Intents, System Logs and UNIX Socket communication. \textit{9} covert channels are implemented which viz. Single and Multiple Settings, Types of Intents, Automatic Intents, Threads Enumeration, UNIX Socket discovery, Free Space on file system, Reading /proc/stat, Timing Channels and Processor Frequency. Experiments are performed to calculate throughput (rate of transmitting data through a channel), stealth (difficulty in identifying a channel), bit-error rate (error-occurance rate in transmission) and required synchronization (timing constraint between source and sink). In tests \textit{4}, \textit{8} and \textit{135} byte data is transmitted from source to sink. During transmission, if the channel under consideration is found open, the information is logged. For the channels which requires synchronization between source and sink, a synchronization protocol is applied on them. This protocol reduces noise and starts measurement on both the ends at the same time. 
\par This paper also proposes a collusion detection approach named \textit{Black-Box} analysis technique. This technique administers a data monitor between applications on the device. The monitor tracks and stores the data used and transmitted by an app to the colluding app. The paper claims that the existing analysis tools such as TaintDroid and XmanDroid failed to detect most of these channels. TaintDroid detects only \textit{2} out of \textit{5} overt channels and \textit{0} out of \textit{9} covert channels. XmanDroid detects \textit{4} out of \textit{5} overt channels and \textit{6} out of \textit{9} covert channels.\\ \\
\textbf{Limitations:} The approach suffers from the following limitations:
\begin{itemize}
	\item The proposed black-box approach is very preliminary, it misses many communications.
	\item Data monitoring cannot handle obfuscation, reflection and encryption. 
	\item It cannot handle complex string analysis that can be used by any of the channels.
\end{itemize}
\subsection{Comparison among state of art approaches}
\label{comparison}
 Researchers have proposed various approaches for intra and inter-app analysis varying from static~\cite{tan2015securing}\cite{arzt2014flowdroid}, dynamic~\cite{petsas2014rage} to policy enforcement~\cite{enck2011defending}\cite{enck2008mitigating} based techniques. In Section~\ref{interAppAnalysis}, we explained these techniques and Section~\ref{stateOfArt} presents research pieces that rely on these techniques.   
 Table~\ref{comparisontabletechniques} summarizes each proposed approach under different criteria: (1) handled components, (2) handled Intents, (3) examines native code or not, (4) resolves reflection or not, (5) works on which code level, (6) conducts intra or inter app analysis and (7) availability of the tool. We believe this helps the reader to examine all the differences in one glance.
 \par Most of the proposed approaches handle Android components viz. Activities(A), Services(S) and Receivers(R), whereas, Content Providers(C) are not handled by ~\cite{chin2011analyzing, wei2014amandroid, asavoae2016towards, wong2016intellidroid, intersection, bhandari2017poster} as shown in column (1) of the table. These approaches consider only Intents as a medium of communication. To access content providers, unique resource identifier (URI) does not use the Intent. Therefore, Intent specific approaches fails to handle content providers. In particular, there are two approaches \cite{Klieber:2014:ATF:2614628.2614633} and \cite{Hay:2015:DDI:2771783.2771800} that are not handling any components other than activities. In \cite{Klieber:2014:ATF:2614628.2614633}, the authors have mentioned that their approach can be similarly extended for other components whereas \cite{Hay:2015:DDI:2771783.2771800} have built a prototype on activities and it is available commercially as a cloud service. In future the authors of \cite{Hay:2015:DDI:2771783.2771800} may extend their approaches to handle all the other components.
 \par There are broadly two types of Intents viz. Implicit(I) and Explicit(E). All the proposed approaches can handle communication through Intents as they are the most popular medium of communication used in Android as shown in column (2) of the table. Although there is one approach \cite{wong2016intellidroid} that is not considering implicit Intent. The reason narrated by the authors of \cite{wong2016intellidroid} is that they do not want to increase false positives. In case of implicit Intent the target is not fixed. If there are multiple receivers then at the run-time one of the receivers is chosen. 
 \par Column (3) of table 1 presents the capability of proposed approaches to handle native code. Native code refers to the code written in C/C++  and used by Android app libraries for low-level interactions with the underlying Linux kernel. Native code runs directly on the processor and hence not included in Dalvik executable that runs in Dalvik virtual machine. Almost all the approaches convert dex into some intermediate representation (IR) language but native code is not get converted into IR and hence, cannot be handled by many tools. However, Flowdroid~\cite{arzt2014flowdroid} can handle very limited native calls as they defined some explicit rules for common invocation of native calls present in Java. Tools like \cite{Klieber:2014:ATF:2614628.2614633, li2015iccta, collusiveDataleak} leverage Flowdroid for analysis and therefore can handle native calls partially.
 \par The proposed approaches based on the their ability to resolve reflection is depicted in column (4) of the table. Reflection is a language's ability to inspect and dynamically call classes, methods, attributes, etc. at runtime. It is a dynamic phenomenon and hence it is very difficult for any static approach to handle it. Dynamic analysis approaches are needed to capture related runtime behaviour features to resolve reflection. If an API is called through reflection, it is passed as a parameter and hence become invisible for detection tools. Although some static tools like~\cite{li2015iccta,ravitch2014multi, collusiveDataleak} can handle reflection partially meaning if the API calls are string constants, then they may be revealed otherwise if they are called through variable where it is obfuscated or encrypted, these tools cannot resolve such reflected calls.
 \renewcommand{\tabcolsep}{0.25cm}
 \renewcommand{\arraystretch}{0.6}
 \begin{sidewaystable*}[htbp]
	\centering
\begin{tabular}{|N|L|M|M|M|M|M|M|M|}
 \toprule [0.12em]
   & \footnotesize{Proposed Approaches} & \footnotesize{Components}& \footnotesize{Intents}& \footnotesize{Native}& \footnotesize{Reflection}& \footnotesize{Code}
   & \footnotesize{Inter-app} & \footnotesize{Availability} \\ 
   & \footnotesize{} & \footnotesize{Handled}& \footnotesize{Handled}& \footnotesize{Code}& \footnotesize{(4)}& \footnotesize{Level}
   & \footnotesize{Analysis}  & \footnotesize{(7)} \\ 
   & \footnotesize{} & \footnotesize{(1)}& \footnotesize{(2)}& \footnotesize{(3)}& \footnotesize{}& \footnotesize{(5)}
   & \footnotesize{(6)} & \footnotesize{} \\ 
   \midrule [0.12em]
   \multirow{10}{*}{\begin{turn}{-270} {\centering \footnotesize{Static}} \end{turn}}

   & \footnotesize{MR-Droid~\cite{MR-Droid}}& \footnotesize{$\langle$A S R C$\rangle$}& \footnotesize{$\langle$E I$\rangle$}& \footnotesize{No}& \footnotesize{No}& \footnotesize{Java Bytecode}& \footnotesize{Yes} & \footnotesize{-}\\ \cline{2-9} 
   
   & \footnotesize{Detecting Inter-App Information Leakage Paths~\cite{bhandari2017poster}}& \footnotesize{$\langle$A S R -$\rangle$}& \footnotesize{$\langle$E I$\rangle$}& \footnotesize{No}& \footnotesize{No}& \footnotesize{Java Bytecode \& Smali}& \footnotesize{Yes} & \footnotesize{-}\\
    \cline{2-9}
    
& \footnotesize{Towards Automated Android App Collusion Detection~\cite{asavoae2016towards}}& \footnotesize{$\langle$A S R -$\rangle$}& \footnotesize{$\langle$E I$\rangle$}& \footnotesize{No}& \footnotesize{No}& \footnotesize{Smali} & \footnotesize{Yes} & \footnotesize{-}  \\  \cline{2-9}  
   
& \footnotesize{ICC Map~\cite{KElishICCMap15}}& \footnotesize{$\langle$A S R C$\rangle$}& \footnotesize{$\langle$E I$\rangle$}& \footnotesize{No}& \footnotesize{No}& \footnotesize{Jimple/Source code}& \footnotesize{Yes} & \footnotesize{-}\\ \cline{2-9}

& \footnotesize{IccTA~\cite{li2015iccta}}& \footnotesize{$\langle$A S R C$\rangle$}& \footnotesize{$\langle$E I$\rangle$}& \footnotesize{Yes*}& \footnotesize{Yes*}& \footnotesize{Jimple}& \footnotesize{Yes$^+$} & \footnotesize{Open-Source}\\ \cline{2-9}

& \footnotesize{Permission Flow~\cite{SbirleaPermissionFlow15}}& \footnotesize{$\langle$A S R C$\rangle$}& \footnotesize{$\langle$E I$\rangle$}& \footnotesize{No}& \footnotesize{No}& \footnotesize{Java Bytecode}& \footnotesize{No} & \footnotesize{-}\\ \cline{2-9}

& \footnotesize{FUSE\cite{ravitch2014multi}}& \footnotesize{$\langle$A S R C$\rangle$}& \footnotesize{$\langle$E I$\rangle$}& \footnotesize{No}& \footnotesize{Yes*}& \footnotesize{Java Bytecode} &\footnotesize{Yes} & \footnotesize{Commercial}  \\\cline{2-9}

& \footnotesize{AmanDroid~\cite{wei2014amandroid}}& \footnotesize{$\langle$A S R - $\rangle$}& \footnotesize{$\langle$E I$\rangle$}& \footnotesize{No}& \footnotesize{No}& \footnotesize{Java Bytecode}& \footnotesize{No} & \footnotesize{Open-Source} \\ \cline{2-9}

& \footnotesize{DidFail~\cite{Klieber:2014:ATF:2614628.2614633}}& \footnotesize{$\langle$A - - -$\rangle$}& \footnotesize{$\langle$E I$\rangle$}& \footnotesize{Yes*}& \footnotesize{No}& \footnotesize{Java Bytecode} & \footnotesize{Yes} & \footnotesize{Open-Source} \\ \cline{2-9}


& \footnotesize{ComDroid~\cite{chin2011analyzing}}& \footnotesize{$\langle$A S R - $\rangle$}& \footnotesize{$\langle$E I$\rangle$}& \footnotesize{No}& \footnotesize{No}& \footnotesize{Java Bytecode}& \footnotesize{No}  & \footnotesize{Open-Source}\\ \hline

\multirow{3}{*}{\begin{turn}{-270} \footnotesize{Dynamic} \end{turn}} 
& \footnotesize{IntelliDroid~\cite{wong2016intellidroid}}& \footnotesize{$\langle$A S R -$\rangle$}& \footnotesize{$\langle$E -$\rangle$}& \footnotesize{Yes}& \footnotesize{Yes}& \footnotesize{Java Bytecode}& \footnotesize{No} & \footnotesize{-}\\
 \cline{2-9}

& \footnotesize{IntentDroid~\cite{hay2015dynamic}} & \footnotesize{$\langle$A - - -$\rangle$}& \footnotesize{$\langle$E I$\rangle$}& \footnotesize{Yes}& \footnotesize{Yes}& \footnotesize{Java Bytecode}& \footnotesize{Yes} & \footnotesize{Commercial}\\ \cline{2-9}
 
 & \footnotesize{TaintDroid~\cite{enck2014taintdroid}}& \footnotesize{$\langle$A S R C$\rangle$}& \footnotesize{$\langle$E I$\rangle$}& \footnotesize{Yes}& \footnotesize{Yes}& \footnotesize{Java Bytecode}& \footnotesize{No} & \footnotesize{Open-Source}\\ \hline
 
 \multirow{4}{*}{\rotatebox{90}{%
  \footnotesize{Policy Enforcement}
}~} 
   & \footnotesize{DIALDroid~\cite{collusiveDataleak}}& \footnotesize{$\langle$A S R C$\rangle$}& \footnotesize{$\langle$E I$\rangle$}& \footnotesize{Yes*}& \footnotesize{Yes*}& \footnotesize{Java Bytecode} & \footnotesize{Yes} & \footnotesize{Open-Source} \\  \cline{2-9}
 & \footnotesize{Intersection Automata Based Model for Android Application Collusion~\cite{intersection}}& \footnotesize{$\langle$A S R -$\rangle$}& \footnotesize{$\langle$E I$\rangle$}& \footnotesize{No}& \footnotesize{No}& \footnotesize{Java Bytecode} & \footnotesize{Yes} & \footnotesize{-} \\  \cline{2-9}
  & \footnotesize{FlaskDroid~\cite{flaskdroid}}& \footnotesize{$\langle$A S R C$\rangle$}& \footnotesize{$\langle$E I$\rangle$}& \footnotesize{No}& \footnotesize{No}& \footnotesize{-} & \footnotesize{Yes} & \footnotesize{-} \\
 \cline{2-9}
 & \footnotesize{XManDroid~\cite{bugiel2011xmandroid}}& \footnotesize{$\langle$A S R C$\rangle$}& \footnotesize{$\langle$E I$\rangle$}& \footnotesize{Yes}& \footnotesize{Yes}& \footnotesize{-}& \footnotesize{Yes} & \footnotesize{-}\\ 
  
\bottomrule [0.12em]
\end{tabular}
\begin{itemize}
\setlength\itemsep{-0.5em}
\footnotesize{
\item A: Activity, S: Service, R: Broadcast Receiver, C: Content Provider
\item E: Explicit Intent, I: Implicit Intent
\item Yes*: The details are explained in section \ref{comparison}
\item Yes$^+$: If it is used with APKCombiner}
\end{itemize}
\caption{Comparison among state of the art approaches}
\label{comparisontabletechniques}
\end{sidewaystable*}
 \par Analysis tools based on the used intermediate representation (IR) for analysis are classified in column (5) of the table. Android APK file is converted to some IR prior to the analysis. There are four code levels on which analysis can be performed viz. Java source code, Java bytecode, Jimple and Smali. Java source code can be analyzed because applications are written in Java language. However, souce is available only if the apps are open-sourced or self developed. Android apps are compiled into Dalvik bytecode called Dex, which is executed in Dalvik virtual machine. For analysis Dalvik should be converted to Java bytecode. This can be done by many APK to Jar converters like dex2jar~\cite{dex2jar}, ded~\cite{octeau2010ded} and Dare~\cite{octeau2012retargeting}. Jimple is a simplified version of Java bytecode. It is a typed 3-address intermediate representation. It is used by Soot~\cite{lam2011soot} which is a popular static analysis framework for Java. Dexpler~\cite{bartel2012dexpler} is a plugin for the Soot framework that translates Dalvik bytecode to Jimple. Smali is another IR used by very popular reverse engineering tool developed by Google named Apktool~\cite{apktool}. Figure~\ref{usedir} shows that Java bytecode is used by most of the approaches, as Java source is generally not available and Jimple and Smali are tool specific representations.
 \begin{figure}
 \centering
 	\includegraphics[scale=0.45]{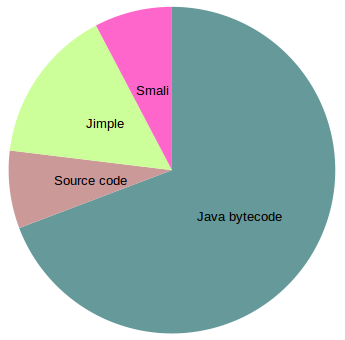}
 	\caption{Intermediate Representation (IR) used for analysis}
 	\label{usedir}
 \end{figure}
\par Apart from this, Figure~\ref{market}, shows the distribution of different app repositories used by the state of art approaches.
\begin{figure}
	\centering
	\includegraphics[scale=0.65]{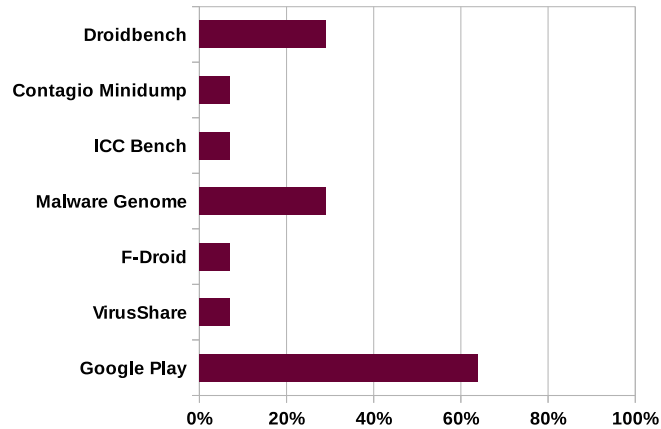}
	\caption{App repositories used for analysis}
	\label{market}
\end{figure}
 \par This brings us to the following conclusions:
\begin{itemize}
\item Intents are considered to be most commonly used ICC communication channels.  Therefore data sharing through Content Providers may become an attractive target for attacker to exploit.
\item Native Calls and Reflection are handled by very few researchers and therefore can be used by malware developers to evade the maliciousness of their code.
\item There are two ways to register Broadcast Receiver viz. static and dynamic. If any static approach that cannot handle reflection but can handle dynamic Broadcast Receiver only works if, registration of Receiver is not reflected.
\item Most of the approaches work on bytecode and therefore maliciousness posed by native code remains untouched.
\item Approaches that analyzes only single app cannot detect inter app leakages.
\end{itemize}
\subsection{Lessons Learned}
\par In this subsection, we mention some recommendations  on which future research needs to focus to stay ahead of smart malware or vulnerabilities. 
\par Analysis approaches should be combination of static, dynamic and policy based analysis techniques to overcome the limitations of all. Other communication channels like Content Providers, Shared Preferences, AIDL etc. should be considered while analysis. Analysis methods should cover different code forms to compensate the losses posed by converting dex to any IR language. Also, analysis methods should consider native code level. There is an urgent need to develop engines that can resolve reflected calls beforehand, so that they are examined during analysis. To capture compositional vulnerabilities like collusion, analysis approaches should examine multiple apps simultaneously. Many proposed approaches are either commercial or not available for public use. It is very difficult for the researchers to evaluate and compare their findings and results. Therefore, it is highly recommended that the tool implemented from the proposed approach along with the tested dataset (if self developed) should be available for free.
\par To prevent unintentional collusion i.e. to avoid the situations when some malicious app can exploit the benign app and use it for collusion, developers need to take special care while signing applications with same certificate and keep their certificate private. Developers should also protect the component that is sending sensitive information to the outside world with specific permissions.

\section{Conclusions}
\label{conclusion}
Android is a modern operating system for smartphones with expanding market share. The main security mechanisms of Android are application sandboxing, application signing, and a permission framework to control access to (sensitive) resources. Android's security framework exhibits serious shortcomings: The burden of approving application permissions is delegated to the end-user who in general does not care much about the impact of prompted permissions on his privacy and security. Hence, malware can be installed on end-user devices such as unauthorized sending of text messages or leaking of sensitive data in the background of running games.  With the growing use  of Android and the awareness of its security vulnerabilities, a number of research contributions have led to tools for the intra-app analysis of Android apps. Unfortunately, these state of the art approaches, and the associated tools, have long left out the security flaws that arise across the boundaries of single apps, in the interaction between several apps. 
We provide in this survey a definition of the collusion in Android, the major security risks on Android,  as well as a summary of the main tools for detecting inter and intra app analysis. The collusion attack is worth investigation. 
\par This survey provides a comprehensive assessment of the strengths and shortcomings of state-of-art approaches. It provides a platform to researchers and practitioners towards proposing technique that can analyze multiple apps simultaneously to detect Android app collusion attacks.

\section{Acknowledgments}
This study has been carried out with financial support from the Department of Information Technology, Government of India Project Grant `Security Analysis Framework
for Android Platform' and the French National Research Agency (ANR)of the French State in the frame of the `Investments for the future' Programme IdEx Bordeaux - CPU
(ANR-10-IDEX-03-02).


\balance
\section*{References}
\biboptions{numbers,sort&compress}
\bibliography{collusionSurvey}

\end{document}